\documentclass[11pt]{article}
\usepackage{jheppub}
\usepackage{amsmath,amssymb,amsfonts,amsthm,mathtools,bm}
\usepackage[dvipsnames]{xcolor}
\usepackage{comment}
\usepackage{hyperref}
\usepackage[utf8]{inputenc}
\usepackage[titletoc]{appendix}
\usepackage{tikz}
\usetikzlibrary{shapes,arrows,shadows}
\usepackage[colorinlistoftodos]{todonotes}
\usepackage{xspace}
\usepackage{etoolbox}
\usepackage{listofitems}
\usepackage{subcaption}
\usepackage{tikz-cd}
\usepackage{physics}

\makeatletter
\def\@fpheader{\relax}
\makeatother

\newcommand\be{\begin{equation}}
\newcommand\ee{\end{equation}}
\newcommand\bea{\begin{eqnarray}}
\newcommand\eea{\end{eqnarray}}
\newcommand\ba{\begin{array}}
\newcommand\ea{\end{array}}

\newcommand\bc{\begin{center}}
\newcommand\ec{\end{center}}

\newcommand{\vol}[1]{\mathsf{vol}_{#1}}
\renewcommand\comment[1]{}
\usepackage{blindtext}

\renewcommand\tilde{\widetilde}
\newcommand\capt[1]{\caption{\textsf{#1}}}

\def\IC{\mathbb{C}}

\def\rd{{\mathrm{d}}}
\newcommand{\define}{\mathrel{\overset{\scriptscriptstyle\rm def}=}}

\newtheorem{definition}{Definition}
\newtheorem{thm}{Theorem}

\newtheorem{lemma}{Lemma}

\newtheoremstyle{indented}{3pt}{3pt}{\addtolength{\leftskip}{2.5em}}{}{\bfseries}{.}{.5em}{}

\newcommand\pf[1]{{\small \begin{proof} #1 \end{proof}}}

\theoremstyle{indented}

 \allowdisplaybreaks
 \numberwithin{equation}{section}

\title{\begin{center}Physical Yukawa Couplings\\in Heterotic String Compactifications\end{center}}

\author[a]{Giorgi Butbaia}
\author[b]{\!\!, Dami\'an Mayorga Pe\~na}
\author[c]{\!\!, Justin Tan}
\author[a]{\!\!, \\ Per Berglund}
\author[d]{\!\!, Tristan H\"ubsch}
\author[e]{\!\!, Vishnu Jejjala}
\author[c]{\!\!, Challenger Mishra}

\affiliation[\,a]{Department of Physics and Astronomy, University of New Hampshire, Durham, NH 03824, USA}
\affiliation[\,b]{CAMSGD,
Department of Mathematics, Instituto Superior T\'ecnico, Universidade de Lisboa,
1049-001 Lisboa, Portugal}
\affiliation[\,c]{Department of Computer Science \& Technology, University of Cambridge, Cambridge CB3 0FD, UK}
\affiliation[\,d]{Department of Physics and Astronomy, Howard University, Washington, DC 20059, USA}
\affiliation[\,e]{Mandelstam Institute for Theoretical Physics, School of Physics, NITheCS, and CoE-MaSS,\\
University of the Witwatersrand, Johannesburg, WITS 2050, South Africa}

\emailAdd{Giorgi.Butbaia@unh.edu}
\emailAdd{damian.mayorga.pena@tecnico.ulisboa.pt}
\emailAdd{jt796@cam.ac.uk}
\emailAdd{Per.Berglund@unh.edu}
\emailAdd{thubsch@howard.edu}
\emailAdd{v.jejjala@wits.ac.za}
\emailAdd{cm2099@cam.ac.uk}

\abstract{
One of the challenges of heterotic compactification on a Calabi--Yau threefold is to determine the physical $(\mathbf{27})^3$ Yukawa couplings of the resulting four-dimensional $\mathcal{N}=1$ theory.
In general, the calculation necessitates knowledge of the Ricci-flat metric.
However, in the standard embedding, which references the tangent bundle, we can compute normalized Yukawa couplings from the Weil--Petersson metric on the moduli space of complex structure deformations of the Calabi--Yau manifold.
In various examples (the Fermat quintic, the intersection of two cubics in $\mathbb{P}^5$, and the Tian--Yau manifold), we calculate the normalized Yukawa couplings for $(2,1)$-forms using the Weil--Petersson metric obtained from the Kodaira--Spencer map.
In cases where $h^{1,1}=1$, this is compared to a complementary calculation based on performing period integrals.
A third expression for the normalized Yukawa couplings is obtained from a machine learned approximate Ricci-flat metric making use of explicit harmonic representatives.
The excellent agreement between the different approaches  opens the door to precision string phenomenology.
}

\usetikzlibrary{fit,shapes,positioning}

\begin{document}

\maketitle
\parskip=5pt

\section{Introduction}
String theory on Calabi--Yau manifolds has offered the promise of deriving the complete structure of the Standard Model of particle physics from the compactification geometry~\cite{Candelas:1985en}.
We focus here on the case of the ``standard embedding''~\cite{Candelas:1985en,Dixon:1989fj,rBeast,candelas2008triadophilia,Candelas:2008wb,Braun:2009qy} of heterotic $E_8\times E_8$ superstring theory on Calabi--Yau threefolds for which there are $\frac12 |\chi|$ generations of particles in the low-energy spectrum.
A modern approach to model building, which invokes bundle structures, does not insist that the Euler characteristic $\chi=\pm 6$ and is perhaps physically more appealing for string phenomenology as we can work with manifolds with a small number of moduli~\cite{Bouchard:2005ag,Braun:2005nv,Douglas:2015aga,Anderson:2013xka,Constantin:2018xkj}. For reviews on the matter see, \textit{e.g.},~\cite{Anderson:2018pui,He:2018jtw}.

Viewed in this more general framework, the  standard embedding is the case in which the vector bundle  $V$ is taken to be the tangent bundle of the Calabi--Yau compatification $T_X$, \textit{i.e.}, the holomorphic sheaf of vector fields on $X$ whose connection solves the Hermitian Yang--Mills equations.  
The standard embedding provides fertile ground for study and is particularly amenable to numerical analysis.
We start here as an initial step. 

By virtue of the Calabi--Yau manifold admitting a Ricci-flat metric in each K\"ahler class, compactification of the heterotic theory on such a geometry preserves $\mathcal{N}=1$ supersymmetry in the four-dimensional effective field theory. %
The Calabi--Yau threefold $X$ is as well an $SU(3)$ holonomy manifold. 
The commutant of $SU(3)$ in $E_8$ is $E_6$, which embeds generations of the particle spectrum of the Standard Model in its $\mathbf{27}$ and $\overline{\mathbf{27}}$ representations.
An important question then is to determine the Yukawa couplings that describe how strongly the low-energy fields interact.
The $(\overline{\mathbf{27}})^3$ couplings are topological (without the worldsheet instantons included), whereas the $(\mathbf{27})^3$ couplings require knowledge of the complex geometry of the Calabi--Yau space~\cite{Strominger:1985ks, CANDELAS1988458,Greene:1987xh}.
In this paper, we focus on the latter set of couplings, expressed in terms of forms resident in $H^{2,1}(X)$, to which we now turn. (For more details, see~\cite{rBeast}.)

The Calabi--Yau condition is also equivalent to the existence of a nowhere vanishing holomorphic top form
\begin{equation}
\Omega^{(3,0)} = \frac{1}{3!} \Omega_{\mu\nu\rho}\, \rd z^\mu \wedge \rd z^\nu \wedge \rd z^\rho \,,
\end{equation}
which provides for the isomorphism $H^{2,1}(X)\overset{\scriptscriptstyle\overline\Omega}\approx H^1(X,T_X)$.
For a given $(2,1)$-form, we can write an equivalent $(0,1)$ $T_X$-valued form as
\begin{equation}
\frac{1}{2!} \omega_{\mu\nu\overline\sigma}\, \rd z^\mu\wedge \rd z^\nu\wedge \rd z^{\overline\sigma} \quad \longleftrightarrow \quad \rd z^{\overline\sigma}(\omega_{\overline\sigma}{}^\mu = \overline{\Omega}^{\mu\nu\rho} \omega_{\nu\rho\overline{\sigma}} )\partial_\mu \,.
\end{equation}
Schematically, the $(\mathbf{27})^3$ couplings are
\begin{equation}
\kappa = \int_X \Omega \wedge \omega^\mu \wedge \omega^\nu \wedge \omega^\rho \, \Omega_{\mu\nu\rho} \,.
 \label{lambda}
\end{equation}
These are the \textit{unnormalized} Yukawa couplings:
the integral depends only on the cohomology class of $\omega$ and not the actual representative.
The \textit{normalized} Yukawa couplings, corresponding to the physical couplings of the model, demand a diagonalization of the kinetic terms.
In general, calculating the normalized couplings requires using the Ricci-flat Calabi--Yau metric in order to choose particular harmonic representatives.

This complication is circumvented in the standard embedding $V \cong T_X$.
In this case, the deformations of $V$ correspond one-to-one to the complex structure deformations of the base Calabi--Yau manifold, and the metric on the $h^{2,1}$-dimensional space of deformations is the Weil--Petersson metric.
The crucial fact is that the Weil--Petersson metric on the moduli space can be calculated without recourse to the Ricci-flat Calabi--Yau metric~\cite{Candelas:1990pi}.
This is sufficient to calculate the normalized Yukawa couplings.

This effort is part of a program to improve numerical results in string phenomenology.
These developments have been revived because machine learning provides good approximations for Ricci-flat Calabi--Yau metrics~\cite{Ashmore:2019wzb,Anderson:2020hux,Douglas:2020hpv,Jejjala:2020wcc,douglas2021holomorphic,Larfors:2022nep,Berglund:2022gvm,Gerdes:2022nzr}
and more recently facilitates the computation of the spectrum of harmonic forms~\cite{Ashmore:2020ujw,Ashmore:2021qdf,Ashmore:2023ajy,Ahmed:2023cnw} that enter into the calculation of Yukawa couplings.
In this work, we use numerical integration techniques to compute the field normalizations and the normalized Yukawa couplings for various heterotic compactifications in the standard embedding.
Furthermore, we use a similar implementation as the one underlying the spectral neural network construction~\cite{Berglund:2022gvm} in order to obtain harmonic, tangent bundle valued $(0,1)$-forms.
The explicit computation of the normalization for those objects requires the Ricci-flat Calabi--Yau metric.
Our aim is to demonstrate that those normalization agree with the Weil--Petersson results. 

We consider two one-parameter Calabi--Yau threefolds, the intersection of two cubics in $\mathbb{P}^5$, the quintic hypersurface in $\mathbb{P}^4$ (and their mirrors) as well as the complete intersection Tian--Yau manifold.
We shall compute the Weil--Petersson metric in two different ways: (i) via a Kodaira--Spencer map for all cases considered, and (ii) via the calculation of period integrals for Calabi--Yau spaces with $h^{2,1}=1$ only.\footnote{
When $h^{2,1}>1$, the analogous calculation requires solving Picard--Fuchs partial differential equations and is more complicated than the $h^{2,1}=1$ case, where we essentially have an ordinary differential equation~\cite{Morrison:1991cd}.}
In particular, we check that for the examples with $h^{2,1}=1$ the Kodaira--Spencer and the period computations agree.
This validates the Kodaira--Spencer algorithms implemented.
These calculations are compared to each other and found to match the canonical computation, using the Ricci-flat Calabi--Yau metric that is calculated numerically using machine learning.

Our work presents the first calculation of physically normalized Yukawa couplings for several complete intersection Calabi--Yau (CICY) manifolds with standard embedding, as well as a versatile extension of the Kodaira--Spencer method advanced by Keller--Lukic~\cite{keller2009numerical}.
The techniques are: firstly, more efficient than machine learning based methods, and secondly, more general than existing methods for computing analytically continued periods, which are anyway only feasible for a small number of moduli.

The organization of the paper is as follows.
In Section~\ref{sec:het}, we sketch the general computation of Yukawa couplings associated to $H^{2,1}(X)$.
Recalling that tangent bundle valued $(0,1)$-forms are dual to $(2,1)$-forms and that these span the massless degrees of freedom transforming as $\mathbf{27}$ under $E_6$, we construct polynomial representatives for the $(0,1)$-forms.
In Section~\ref{sec:KSmap}, we discuss the Kodaira--Spencer map and its use in the computation of the field normalizations (see Section~\ref{ssec:KS-WP}).
In Section~\ref{sec:numerical}, we present numerical results.
For the mirror of the intersection of two cubics in $\mathbb{P}^5$, we compare the period integral result with the numerical integration that produces the Weil--Petersson metric and demonstrate the agreement of both methods.
We also compute the Yukawa coupling for this example and show that it agrees with the period result for any value of the modular deformation.
In addition, we consider a $\mathbb{Z}_5\times\mathbb{Z}_5$ quotient of the Fermat quintic and the $\mathbb{Z}_3$ quotient of the Tian--Yau manifold.
For the Fermat quotient, we compare the normalized Yukawa couplings to the conformal field theory computations~\cite{DISTLER1988295} and show that the Kodaira--Spencer normalization produces the correct results.
Similarly, for the Tian--Yau quotient, we contrast our computation with the unnormalized Yukawa couplings of~\cite{CANDELAS1988357}.
For this case, we obtain the normalized couplings as well as their behavior along a modulus direction.
Our methods are general and can be readily applied to the standard embedding of any complete intersection Calabi--Yau manifold.
In Section~\ref{sec:mlh}, we discuss the method employed to search for the harmonic representatives and the direct computation of the normalizations which makes use of machine learned Ricci-flat metric. We discuss our implementation for the quintic and the bicubic.
Finally, in Section~\ref{sec:discussion}, we present conclusions and prospects for future work.

The numerical implementation of the Weil--Petersson metric as well as the calculation of approximate Ricci-flat Calabi--Yau metrics are part of a \textsf{JAX}~\cite{jax2018github} library called \textsf{cymyc} (\underline{C}alabi--\underline{Y}au \underline{M}etrics, \underline{Y}ukawas, and \underline{C}urvature)~\cite{cymyc}, to be released soon.

\textbf{Note added:}
One week after the first version of this preprint appeared on arXiv, the preprint~\cite{Constantin:2024yxh} considered machine learning based calculation of physical Yukawa couplings in non-standard embeddings.

\section{Heterotic Yukawa couplings}\label{sec:het}
Let us begin by considering the $E_8 \times E_8$ heterotic string compactified on a Calabi--Yau threefold $X$.
The compactification breaks $E_8 \times E_8$ to a smaller subgroup.
In order for $\mathcal{N}=1$ supersymmetry to be preserved in the resulting four-dimensional effective theory, the structure group $H$ of a principal bundle $V$ over $X$ must be embedded into $E_8\times E_8$.
The matter in four dimensions can be obtained from the corresponding decomposition of the $E_8\times E_8$ adjoint representation.

For simplicity, consider a subgroup  $G$ in a single $E_8$ with $G$ the commutant of $H$ in $E_8$, so that the effective gauge symmetry in four dimensions is $G\times E_8$.
The matter in the visible sector is then supplied by the decomposition of the $248$-dimensional adjoint representation of $E_8$,
\begin{equation}\label{248->...}
    \mathbf{248}=\mathbf{Adj}_{E_8}\rightarrow
    (\mathbf{Adj}_{H},\mathbf{1}) ~\oplus~ (\mathbf{1},\mathbf{Adj}_{G})
    ~\bigoplus_i (\mathbf{R}_H^{(i)},\mathbf{R}_G^{(i)}) \,,
\end{equation}
where $\mathbf{R}_H^{(i)}$ and $\mathbf{R}_G^{(i)}$ are suitable representations of $H$ and $G$.
More specifically, matter in the representation $\mathbf{R}_G^{(i)}$ of the effective gauge group is represented by harmonic $(0,1)$-forms $a^{(i)}$ that take values in a vector bundle $V_i$,\footnote{The index $i$ denotes different bundles, such as $V$, $V^*$, and $V\otimes V^*$.} \textit{i.e.}, $a^{(i)}\in H^1(X,V_i)$.

We discuss computation of the trilinear interaction terms.
The holomorphic Yukawa couplings $\lambda(a^{(i)},b^{(j)},c^{(k)})$ may be nonzero provided the tensor product $\mathbf{R}_G^{(i)}{\otimes}\mathbf{R}_G^{(j)}{\otimes}\mathbf{R}_G^{(k)}$ contains a $G$-invariant, and may be computed, generalizing~\eqref{lambda}, as 
\begin{equation}\label{eq:yuk1}
    \kappa(a^{(i)},b^{(j)},c^{(k)})=\int_X \Omega \wedge
    \widetilde\Omega \big(a^{(i)},b^{(j)},c^{(k)}\big)\,, 
\end{equation} 
where $\Omega$ is the holomorphic $(3,0)$ form and $\widetilde\Omega (a^{(i)},b^{(j)},c^{(k)})$ is the appropriate contraction with the $(0,1) \, V$-valued forms, with $\widetilde\Omega$ a suitable deformation of the standard $\Omega$ as long as $V_i$ is a rank-$3$ deformation of $T_X$.
The couplings~\eqref{eq:yuk1} only become the physical ones once we know the K\"ahler potential for the matter fields, which yields the corresponding kinetic terms.

The low-energy effective action of an $\mathcal{N}=1$ theory is written as
\begin{equation}
S_\text{eff} = \int \rd^4x\ \left[ \int \rd^4\theta\ K\big(\Phi^a, \overline{\Phi}{}^{\overline{b}}\big) + \frac{1}{4g^2} \left( \int \rd^2\theta\ \text{tr}\, \mathcal{W}_\alpha \mathcal{W}^\alpha + \int \rd^2\theta\ W(\Phi^a) \right) + \text{h.c.} \right], \label{eq:action}
\end{equation}
where $\Phi^a$ are chiral superfields and $\mathcal{W}_\alpha$ is the gauge field strength associated to the vector superfield.
The superpotential $W(\Phi^a)$ is a holomorphic function of the superfields and is gauge invariant with $R$-charge $2$.
The Yukawa couplings originate from this term in the effective action.
The K\"ahler potential, which is explicitly not holomorphic, contains the kinetic terms:\footnote{From the underlying worldsheet quantum field theory, this kinetic term normalization metric emerges as a two-point correlation function defining (the appropriate generalization of) the Zamolodchikov metric~\cite{Zamolodchikov:1986gt}. In the special case when $V_i=T_X$, this equals the Weil--Petersson metric~\cite{Candelas:1989qn}.}

\begin{equation}
K\big(\Phi^a, \overline{\Phi}{}^{\overline{b}}\big) \supset N_{a\overline{b}} \; \Phi^a \overline{\Phi}{}^{\overline{b}} + \ldots \,.
\end{equation}
The entries of the normalization matrix $N_{a\overline{b}}$ are proportional to the inner product 
\begin{equation}\label{eq:norm} 
    N_{a\overline{b}}\sim (a,b)=\int_X a \wedge \bar{\star}_V b\,,
\end{equation}
between the harmonic representatives $a,b$ of their respective classes in $H^1(X,V_{i})$.
Equipped with this inner product, starting from a given basis $\{a_k^{(i)}\}_{k=1}^{h^1(V_i)}$, we may obtain an orthonormal basis $\{a_k^{(i)\prime}\}_{k=1}^{h^1(V_i)}$ via diagonalization of the normalization matrix induced by (\ref{eq:norm}) and rescaling by the square root of the eigenvalues, \textit{i.e.}, the normalizations  $N_{a^\prime_k}$ of each eigenform $a_k^{(i)\prime}$. This change of basis converts the holomorphic Yukawa couplings into the physical Yukawa couplings, computed as 
\begin{equation}
   Y\big(a^{(i)\,\prime}, b^{(j)\,\prime}, c^{(k)\,\prime}\big)
   =\frac{\int_X \Omega \wedge \tilde{\Omega}\big(a^{(i)\,\prime}, b^{(j)\,\prime}, c^{(k)\,\prime}\big)}{\int_X \Omega\wedge\overline{\Omega}}\,.
 \label{eq:tripleYuk}
\end{equation}

Reflecting on what we have discussed so far, we emphasize the following two points.
\begin{itemize}
\item The computation of the holomorphic Yukawa couplings in~\eqref{eq:yuk1} does not require knowledge of the harmonic representatives in $H^1(X, V_i)$, \textit{i.e.}, the unnormalized couplings are the same when computed using elements in the cohomology classes $[a^{(i)}]$,  $[b^{(j)}]$, and $[c^{(k)}]$.
The calculation of $\lambda$ is {\em quasi-topological\/}%
~\cite{Blesneag:2015pvz, Blesneag:2016yag}.
\item 
This is not the case for the normalization~\eqref{eq:norm}, since the Hodge star $\star_V$  between harmonic bundle-valued forms requires knowledge of the Ricci-flat metric on $X$ and the Hermitian structure on $V$.
The calculation of $\kappa$ requires geometric input.
\end{itemize}

Here we briefly note that one requires geometric computations involving the metric to utilize the Ricci-flat representative for the Kähler class being considered in order for the metric on the Calabi--Yau complex structure moduli space to be Kähler, an argument we will make more precise in the discussion after Lemma~\ref{lemma:WPHarmonic}.

Let us now specialize to the case where $H=SU(3)$, for which the adjoint decomposition takes the form 
\begin{equation}
    \mathbf{248}\rightarrow (\mathbf{78},\mathbf{1})\oplus (\mathbf{1},\mathbf{8})\oplus  (\overline{\mathbf{27}},\overline{\mathbf{3}})\oplus  (\mathbf{27},\mathbf{3})\,.
\end{equation}
As $E_6 \times SU(3)$ is a maximal subgroup of $E_8$, we identify $E_6$ as the GUT gauge group $G$.
The number of $\mathbf{27}$ multiplets are counted by $h^1(V)$ while the number of $\overline{\mathbf{27}}$ multiplets are counted by $h^1(V^*)=h^2(V)$.
There might also be additional singlet fields corresponding to bundle moduli; these are counted by $h^1(V\otimes V^*)$.
In the standard embedding, the role of the holomorphic vector bundle $V$ is played by the tangent sheaf $T_X$, whose structure group is indeed $SU(3)$.
Setting $V=T_X$ implies that the difference between the number of massless $\mathbf{27}$ and $\overline{\mathbf{27}}$ representations is an index, half of the Euler characteristic. It also motivates the following lemma (see Appendix~\ref{sec:proofs} for a detailed proof):
\begin{lemma}\label{lemma:Serre} 
\addtolength{\leftskip}{\parindent}
Let $X$ be a Calabi--Yau manifold, then: $H^1(X,T_X)\simeq H^{n-1,1}_{\overline{\partial}}(X)$ and the isomorphism is given by:
\begin{gather}
	[\alpha] \longmapsto [\Omega(\alpha)]\,,
 \label{e:Serre}
\end{gather}
where $\Omega\in H^{n,0}_{\overline{\partial}}(X)$ is nowhere zero.
\end{lemma}

For the particular case of Calabi--Yau threefolds, this implies the well-known isomorphism $H^1(X,T_X) \simeq H^{2,1}(X)$. Recall further that for this particular case, the pairing~\eqref{eq:norm} becomes the Weil--Petersson metric on the Calabi--Yau complex structure moduli space (see also Definition~\ref{defWP} below),
\begin{equation}\label{eq:wp_metric}
    (a,b)=\int_X a\wedge \bar{\star}_g b \define\langle a,b \rangle_{\rm WP}\,.
\end{equation}

This may be computed by exploiting the existence of the Ricci-flat metric without its direct invocation, owing to special geometry, as we shall see in the sequel.

We are interested in the computation of Yukawa couplings of the form $(\mathbf{27})^3$ which involve only elements in $H^1(X,T_X)$. In this case, the pairing introduced in~\eqref{eq:yuk1} can be written as 
\begin{equation}
    \widetilde{\Omega}(a,b,c) \define
     a^\mu \wedge b^\nu \wedge c^\rho\, \Omega_{\mu\nu\rho}\,,
\end{equation}
where the $(3,0)$-form $\Omega$ acts by contraction on the $\bigwedge^3 T_X$-valued $(0,3)$-form to give an ordinary, $\IC$-valued $(0,3)$-form. If we take an orthogonal basis $\{a_k\}_{k=1}^{h^1(T_X)}$ the corresponding normalized Yukawa couplings take the form 
\begin{gather}\label{eq:NormYuk}
	Y_{ijk}
 =\frac{\displaystyle \int_{X} \Omega\wedge\widetilde{\Omega}(a_i, a_j, a_k)}
 {\displaystyle\sqrt{N_{i} N_{j} N_{k}}\int_{X} \Omega\wedge\overline{\Omega}}\,.
\end{gather}

\section{Physical Yukawa couplings via the Kodaira--Spencer map}
\label{sec:KSmap}
\subsection{Computing normalizations}
\label{ssec:KS-WP}
In order to discuss the computation of the canonical normalization matrix via the Weil--Petersson metric (\ref{eq:wp_metric}), we briefly recall some facts on the metric on the complex structure moduli of a Calabi--Yau manifold $X$ with K\"ahler class $[\omega]\in H_{\overline{\partial}}^{1,1}(X)$. We shall mostly follow the notation of~\cite{tian:1987}. Let us start by considering a complex analytic family (in the sense of Kodaira and Spencer~\cite{56511be9-71c1-3ccb-89df-a73bcdcc07ed}, for more details see:~\cite{kodaira_2005}) $(\mathcal{X}, B,\varpi)$ of Calabi--Yau manifolds over a base $B$ such that $0\in B \subseteq \IC^{\dim H^1(X, T_X)}$ with projection map $\varpi\colon \mathcal{X}\rightarrow B$ and $X_t := \varpi^{-1}(t)$ with  $X := X_0$. Recall that the Weil--Petersson metric $\langle-,-\rangle_{\mathrm{WP}}$ on the moduli space $T_0B$ of complex deformations of $X$ can be written as a K\"ahler metric such that $X_0 = X$ and $[\omega_t] = [\omega]$. We shall refer to this family as a polarized complex analytic family, with polarization induced by $[\omega]$.

\begin{definition} \label{defWP}
\addtolength{\leftskip}{\parindent}
Let $g_t \in [\omega]$ denote the unique Ricci-flat metric on $X_t \in \mathcal{X}$ in the polarized complex analytic family $(\mathcal{X}, B,\varpi)$. The Weil--Petersson metric is then defined as:
\begin{gather}\label{eq:defWP}
	\langle a, b\rangle_{\mathrm{WP}} = \int_{X_t}\rho(a)\wedge \overline{\star_{g_t}}\mathcal{H}\rho(b)\,,
\end{gather}
where $\rho\colon T_t B\rightarrow H^1(X,T_X)$ is the Kodaira--Spencer map~\cite{56511be9-71c1-3ccb-89df-a73bcdcc07ed, kodaira_2005} 
and $\mathcal{H}\colon H^{p,q}_{\overline{\partial}}(X) \rightarrow \mathcal{H}^{p,q}_{\overline{\partial}}(X)$ is the harmonic projection.
\end{definition}

The Kodaira--Spencer map can be defined in the following manner: Let $\{U_j\}$ be a finite cover of $X_t$ with local coordinates $\{z_j^1, \dots, z_j^n\}$ such that for every $U_{j}\cap U_{k}\neq \emptyset$ the gluing maps are given by $f_{jk}\colon B\times U_{k} \rightarrow U_{j}$. Then, the Kodaira--Spencer map is defined as follows:
\begin{gather}
    \rho\left(\frac{\partial}{\partial t}\right) = \left[\left\{\frac{\partial f_{jk}^\mu(z_k, t)}{\partial t}\frac{\partial}{\partial z^\mu_{j}}\right\}\right],\quad\text{where}~z_k = f_{kj}(z_j,t) \,.
\end{gather}

Note that from~\cite{tian:1987} we have $\mathrm{im}\, {\rho} \subseteq H^1(X,T_X)_{\omega}$ where $H^1(X,T_X)_{\omega}$ is the subspace of polarization preserving deformations: $[\phi] \in H^1(X,T_X)_{\omega}$ if $[\omega(\phi)] = 0$. It can be shown that if $\phi \in [\phi]$ is harmonic, then $g_t(\phi) = 0$ identically~\cite{nannicini:1986}. This leads to the following lemma:
\begin{lemma}\label{lemma:WPHarmonic}
\addtolength{\leftskip}{\parindent}
Let $\Omega \in H_{\overline{\partial}}^{n,0}(X_t)$ where $n=\dim{X_t}$, be non-zero, then:
\begin{gather}
	\langle a, b\rangle_{\mathrm{WP}} = - \frac{\displaystyle \int_{X_t} \Omega(\mathcal{H}\rho(a)) \wedge \overline{\Omega(\mathcal{H}\rho(b))}}{\displaystyle \int_{X_t} \Omega \wedge \overline{\Omega}} \int_{X_t}\vol{g_t}\,.
 \label{L:WPH}
\end{gather}
\end{lemma}
\pf{\addtolength{\leftskip}{\parindent}
See Appendix~\ref{sec:proofs} or Refs.~\cite{todorov:1989, tian:1987}.} At first glance, it may seem that evaluation 
of~\eqref{eq:defWP} requires the metric $g_t$ on $X_t$, which also induces a metric on $\Omega^{0,1}(T_{X_t})$. However, the Ricci-flatness consequence $\textsf{vol}_{g_t} \propto \Omega_t \wedge \overline{\Omega_t}$ ensures~\eqref{eq:norm} may be expressed in terms of the standard cup product on $H^{p,q}_{\bar{\partial}}(X_t)$ with $p+q=n$. 

Since our numerical methods use representatives of the Kodaira--Spencer classes $\rho(a) \in H^1(X,T_X)$ which are not necessarily harmonic, the polarization-preserving condition $g_t(\phi)=0$ is not necessarily guaranteed to hold. We therefore show explicitly, at the level of forms, that the Weil--Petersson metric may be computed with arbitrary representatives. We shall first briefly recall the explicit construction of the Kodaira--Spencer class:

Recall that for any $t,t'\in B$, we have $X_t \simeq X_{t'}$ diffeomorphic as real manifolds. For simplicity, let $t'=0$, and denote the diffeomorphism by:
\begin{gather}
	\zeta_t\colon X \stackrel{\simeq}{\longrightarrow} X_t\,.
\end{gather}
Then, the corresponding infinitesimal deformation $\xi$ to $\zeta_t$ at $t=0$ is a set of non-holomorphic vector fields: $\xi = \{\xi_j\}_j$ defined on a finite open cover $\{U_j\}_j$ of $X$. Then, we may construct Kodaira--Spencer class corresponding to $\xi$ using \v{C}ech co-cycle defined by:
\begin{gather}
	[\{\overline{\partial}\xi_j\}_j] \in \check{H}^1(X,T_X)\simeq H^1(X,T_X)\,.
\end{gather}

From the results of~\cite{tian:1987, todorov:1989}, the $\langle-,-\rangle_{\mathrm{WP}}$ is shown to be a K\"ahler metric with the local K\"ahler potential given by the canonical intersection pairing on $H^{n,0}_{\overline{\partial}}(X)$. We shall show that such identification can also be computed with arbitrary representatives without application of harmonic projections. In particular, let $\Omega(t)$ be a holomorphic $n$-form on the total space which is smoothly varying with respect to the deformation parameter $t\in B$ of the polarized complex analytic family $(\mathcal{X}, B,\varpi)$ and restricts to a non-zero holomorphic $(n,0)$ form $\Omega_t \in H^{n,0}_{\overline{\partial}}(X_t)$ on each fibre. Then one has the following decomposition:
\begin{gather}\label{eq:domegaDecomposition}
	\frac{\rd\Omega_t}{\rd t}\bigg\vert_{t=0} 
 = 	\Omega' + \Omega(\phi) ~\in~
 \Gamma(X, \Omega^{n,0}) \oplus \Gamma(X, \Omega^{n-1,1}), 
 ~~\text{respectively},
\end{gather}
where $\phi \in H^1(T_X)$ is a representative of the Kodaira--Spencer class. The arguments of~\cite{tian:1987, todorov:1989, nannicini:1986} apply the harmonic projection to $\phi$ to show that $\Omega'$ is holomorphic. Since the general numerical methods that we consider do not compute harmonic representatives, we show that the result holds true for an arbitrary choice of the representatives, when the $(n,0)$ part $\Omega'$ of the variation in the canonical holomorphic form is not necessarily holomorphic.
\begin{thm}\label{thm:wpIndependentFromMetric}
\addtolength{\leftskip}{\parindent}
The identification of $\langle -, -\rangle_{\mathrm{WP}}$ with the K\"ahler metric using~\eqref{eq:domegaDecomposition} is true for an arbitrary choice of representatives of the Kodaira--Spencer class.
\end{thm}
\pf{\addtolength{\leftskip}{\parindent}
Let $(-,-)$ denote the intersection pairing on $H^{p,q}_{\overline{\partial}}(X)$ with $p+q=n$:\begin{gather}
	(\alpha,\beta) = \int_X \alpha\wedge \overline{\beta}\,.
\end{gather}
Then, without loss of generality, we shall show that:
\begin{gather}\label{eq:WP_and_domega}
	\frac{\displaystyle \langle a, a\rangle_{\mathrm{WP}}}
         {\displaystyle \int_{X}\vol{g}} ~=~
    \frac{\displaystyle\left(\frac{\rd\Omega_t}{\rd  t}\bigg\vert_{t=0}, \frac{\rd\Omega_t}{\rd t}\bigg\vert_{t=0}\right)}{(\Omega, \Omega)} + \frac{\displaystyle \left|\left(\Omega, \frac{\rd\Omega_t}{\rd t}\bigg\vert_{t=0}\right)\right|^2}{(\Omega, \Omega)^2}\,,
\end{gather}
where the decomposition~\eqref{eq:domegaDecomposition} is arbitrary and $a\in T_0B$ such that $\rho(a) = [\phi]$. This implies that the terms in~\eqref{eq:WP_and_domega} due to the $\Gamma(X, \Omega^{n,0})$ component of the decomposition~\eqref{eq:domegaDecomposition} are not necessarily zero. Recall that closure is a topological condition; we have: $d\Omega_t = 0$ for all $t$, which implies:
\begin{gather}\label{eq:OmegaClosedEquation}
	\partial \Omega' + \overline{\partial}\Omega' + \partial(\Omega(\phi)) = 0\,.
\end{gather}
Note that $\deg{\partial\Omega ' } = (n+1, 0)$ whereas $\deg{\overline{\partial}\Omega'} = \deg \partial(\Omega(\phi)) = (n,1)$, thus $\partial\Omega' = 0$ due to the Hodge decomposition. Let $\psi \in \Gamma(X, \Omega^{n-1,0})$ be such that:
\begin{gather}\label{eq:harmonicPsi}
	\Omega(\phi) + \overline{\partial}\psi = \mathcal{H}\Omega(\phi)\,,
\end{gather}
whose existence is guaranteed by the Hodge theorem. Then, by combining~\eqref{eq:OmegaClosedEquation} and~\eqref{eq:harmonicPsi} we obtain:
\begin{gather}
	\overline{\partial}\Omega' + \partial(\Omega(\phi)) + \partial\overline{\partial}\psi - \overline{\partial}\partial \psi = \overline{\partial}(\Omega' + \partial\psi) + \partial\mathcal{H}(\Omega(\phi)) = 0\,.
\end{gather}
However, using K\"ahler identities, we have: $\partial\mathcal{H}(\Omega(\phi)) = 0$, thus $\Omega' + \partial\psi = c \Omega$ for some constant $c\in \IC$ by compactness. Thus, to show that~\eqref{eq:WP_and_domega} is true, it remains to compute the intersection products. In particular, we have:
\begin{gather}
	\left(\frac{\rd\Omega_t}{\rd t}\bigg\vert_{t=0}, \frac{\rd\Omega_t}{\rd t}\bigg\vert_{t=0}\right) = (\Omega', \Omega') + (\Omega(\phi), \Omega(\phi))\,,
\end{gather}
where the $(\Omega', \Omega')$ can be decomposed as:
\begin{gather}
	(\Omega', \Omega') = (c\Omega - \partial\psi, c\Omega - \partial\psi) = |c|^2(\Omega, \Omega)	 + (\partial\psi, \partial\psi)\,,
\end{gather}
where we have used $\partial\Omega = 0$. Note that $(\partial\psi, \partial \psi)$ is not necessarily zero. Similarly, we may decompose $(\Omega(\phi), \Omega(\phi))$ as:
\begin{gather}
	(\Omega(\phi), \Omega(\phi)) = (\mathcal{H}\Omega(\phi) - \overline{\partial}\psi, \mathcal{H}\Omega(\phi) - \overline{\partial}\psi)	 =  (\mathcal{H}\Omega(\phi), \mathcal{H}\Omega(\phi)) +  ( \overline{\partial}\psi, \overline{\partial}\psi)\,,
\end{gather}
where we have used the harmonicity condition. A simple integration by parts argument and application of Stokes' theorem gives: $(\overline{\partial}\psi, \overline{\partial}\psi) = -(\psi, \partial\overline{\partial}\psi)$ and $(\partial\psi ,\partial\psi) = -(\psi, \overline{\partial}\partial\psi) = (\psi, \partial\overline{\partial}\psi)$, thus, we have:
\begin{gather}
	\left(\frac{\rd\Omega_t}{\rd t}\bigg\vert_{t=0}, \frac{\rd\Omega_t}{\rd t}\bigg\vert_{t=0}\right) = |c|^2(\Omega, \Omega) + \big(\mathcal{H}\Omega(\phi), \mathcal{H}\Omega(\phi)\big)\,.
\end{gather}
Finally, we have:
\begin{gather}
	\left(\frac{\rd\Omega_t}{\rd t}\bigg\vert_{t=0}, \Omega\right) = (\Omega', \Omega) = (c\Omega - \partial\psi, \Omega) 
 =c(\Omega, \Omega)\,.
\end{gather}
From this, direct calculation shows that:
\begin{gather}\label{eq:dOmegaResultHarmonic}
	-\frac{\displaystyle\left(\frac{\rd\Omega_t}{\rd t}\bigg\vert_{t=0}, \frac{\rd\Omega_t}{\rd t}\bigg\vert_{t=0}\right)}{(\Omega, \Omega)} + \frac{\displaystyle \left|\left(\Omega, \frac{\rd\Omega_t}{\rd t}\bigg\vert_{t=0}\right)\right|^2}{(\Omega, \Omega)^2} = -\frac{\big(\mathcal{H}\Omega(\phi), \mathcal{H}\Omega(\phi)\big)}{(\Omega, \Omega)}\,.
\end{gather}
The result then follows from the statement of Lemma~\ref{lemma:WPHarmonic}, where we have used the fact that $\mathcal{H}(\Omega(\phi)) = \Omega(\mathcal{H}\phi)$, which follows from Ricci-flatness of $X_t$~\cite{todorov:1989}.}

\subsection{Constructing the Kodaira--Spencer map}
\label{ssec:constrKSmap}

In this section we shall briefly review the method described in~\cite{keller2009numerical} and show that it naturally generalizes to Calabi--Yau complete intersections. %
The main idea described in~\cite{keller2009numerical} is to find explicit form of the decomposition:~\eqref{eq:domegaDecomposition} and then apply numerical integration techniques to compute the canonical intersection pairings~\eqref{eq:dOmegaResultHarmonic}. As described in Section~\ref{ssec:KS-WP}, the $(n,0)$-terms in the decomposition~\eqref{eq:domegaDecomposition} now contribute non-trivially to the Weil--Petersson metric and both components must be computed explicitly.

First, let us briefly set up the notation defining the complete intersection Calabi--Yau $X$. Recall that the information defining $X$ can be given as a configuration matrix:
\begin{gather}\label{eq:configurationMatrix}
	X = \left[\begin{matrix}
		\begin{matrix}
			n_1 \\
			\vdots \\
			n_m
		\end{matrix}
		~\vline~
		\begin{matrix}
			q_1^1 & \dots & q_K^1 \\
			\vdots & \ddots & \vdots \\
			q_1^N & \dots & q_K^N
		\end{matrix}
	\end{matrix}\right]\,,
\end{gather}
where, after fixing the point in the complex structure moduli,  the defining equations are given by polynomials: $\{p_j\}_{j=1}^K$ of appropriate degrees specified by~\eqref{eq:configurationMatrix}. To describe the deformations of $X$, we may identify $H^1(T_X)$ with a quotient~\cite{Green:1987cr}:
\begin{gather}\label{eq:deformationsAndCohomology}
	H^1(T_X)\simeq \bigoplus_{l=1}^K H^0(X, \mathcal{O}_X(q_l)) \bigg\slash \sim \qquad \ni F\,.
\end{gather}
Thus, suppose $\zeta_t\colon X \longrightarrow X_t$ denotes the diffeomorphism induced by the deformation $F$ in~\eqref{eq:deformationsAndCohomology}, then we have $p_j(\zeta_t) + tF_j(\zeta_t) = 0$ for all $j\in \{1,\dots,K\}$. Let $\{Z^j\}_{j=1}^M$ for $M = n_1 + \dots + n_N$ be the set of local coordinates on the ambient space: $A = \mathbb{P}^{n_1}\times \cdots \times \mathbb{P}^{n_N}$. Let $\xi \in T_X$ be the generator of the diffeomorphism corresponding to $F$. Note that $\xi$ is a non-holomorphic section of $T_A$ that satisfies:
\begin{gather}\label{eq:infinitesimalDiffeomorphism}
	\frac{\partial p_j}{\partial Z^k}\xi^k + F_j = 0,\quad\text{for all}~j\in \{1,\cdots,K\}\,.
\end{gather}
From above it follows that the solution for $\xi$ is not unique. In~\cite{keller2009numerical, CANDELAS1988458} the authors give examples of particular solutions to~\eqref{eq:infinitesimalDiffeomorphism}. We shall instead derive a general solution. Let $\mathrm{Jac}(p)$ denote the matrix $\{\partial p_j/\partial Z^k\}_{j,k}$. Then, after fixing a metric $g$ on the ambient space, it suffices to compute Moore--Penrose pseudoinverse with respect to metric $g$ and kernel of $\mathrm{Jac}(p)$. Explicitly, the right pseudoinverse $\mathrm{Jac}(p)^+$ with respect to $g$ is given by:
\begin{gather}
	\mathrm{Jac}(p)^+ = \overline{\mathrm{Jac}(p)}\cdot\big(\mathrm{Jac}(p)\cdot \overline{\mathrm{Jac}(p)}\big)^{-1}\,,
\end{gather}
where all inner products are induced by the metric $g$. Thus, a general solution to~\eqref{eq:infinitesimalDiffeomorphism} is given by:
\begin{gather}\label{eq:generalSolution}
	\xi^k = (\mathrm{Jac(p)}^+)^{kj} F_j + \sum_{i}c_i \chi^{k}_i,\quad\text{for}~c_i\in\IC\,,
\end{gather}
where $\IC\{ \chi_i\}_i = \ker{\mathrm{Jac}(p)}$. Note that the expressions given in~\cite{keller2009numerical, CANDELAS1988458} correspond to the coefficients $c_i = 0$~\eqref{eq:generalSolution}, but they differ by the choice of the metric $g$ on the ambient space.

From the non-holomorphic local vector field $\xi$ corresponding to the deformation $F$, we may compute the Kodaira--Spencer class using the method described in Section~\ref{ssec:KS-WP} as:
\begin{gather}
	\rho\left(\frac{\partial}{\partial t}\right) = \left[\overline{\partial}\xi\right]\,.
\end{gather}
Which allows us to compute unnormalized Yukawa coupling using the Kodaira--Spencer map in the following manner, let $\xi_{i}$ be the vector field corresponding to $\rho(\partial/\partial t_i)$, then, the unnormalized Yukawa coupling $\kappa_{ijk}$ is given by the following integral:
\begin{gather}
    \kappa_{ijk} = \int_{X} \Omega_{\alpha\beta\gamma}\frac{\partial \xi^\alpha_i}{\partial \overline{z}^\mu}\frac{\partial \xi^\beta_j}{\partial\overline{z}^\nu}\frac{\partial\xi^{\gamma}_k}{\partial \overline{z}^\delta}~\Omega\wedge d\overline{z}^\mu \wedge d\overline{z}^\nu \wedge d\overline{z}^\delta
\end{gather}

Thus, what remains to compute is the decomposition~\eqref{eq:domegaDecomposition}. In the case of hypersurfaces, this has been done in~\cite{keller2009numerical}. We show that this naturally generalizes to arbitrary complete intersections with $\mathrm{codim}>1$. This can be done by differentiating the Poincar\'e residue equation. First, note that for a sufficiently small $t>0$, the vector field $\xi$ in $T_A$ induces an automorphism of the ambient space $A$:
\begin{equation}
	\begin{tikzcd}
		A \arrow{r}{}{\check{\zeta}_t} & A_t \\
		X \arrow{u}{} \arrow{r}{\zeta_t} & X_t \arrow{u}
	\end{tikzcd}
	\label{eq:diag}
\end{equation}
 Let $U_t \subset A_t$ be some open set, and pick local coordinates $(Z_t^1,\dots, Z_t^M)$ on $U_t$, such that $Z_0^j = Z^j$ for all $j$, where $M = \dim{A_t}$. Using the adjunction formula, we may relate canonical bundles of $X_t$ and $A_t$ as: $K_{X_t} = K_{A_t} \otimes \det{\mathcal{N}_{X_t|A_t}}\big\vert_{X_t}$, which leads to:
 \begin{gather}\label{eq:adjunctionFormula}
 	\Omega_t \wedge \rd p^1_t \wedge \cdots \wedge \rd p^K_t = \rd Z_t^1\wedge \cdots \wedge \rd Z_t^M\,,
 \end{gather}
where $\Omega_t \in \Gamma(X_t, K_{X_t})$. Following~\cite{keller2009numerical}, we consider perturbation of~\eqref{eq:adjunctionFormula} at $t=0$ in the direction $\xi \in T_A$. In particular, we have:
\begin{gather}
	\frac{\rd}{\rd t}\left(\rd Z_t^1 \wedge \cdots \wedge \rd Z_t^M\right)\bigg\vert_{t=0} = \frac{\rd}{\rd t}\det\left(\delta_k^j + t \frac{\partial \xi^j}{\partial Z^k}\right)_{j,k}\bigg\vert_{t=0}\rd Z^1 \wedge \cdots \wedge \rd Z^M + \\ \notag
	+ \sum_{j,k=1}^M\frac{\partial \xi^j}{\partial \overline{Z}^k}\rd Z^1\wedge \cdots \wedge\widehat{\rd Z^j}\wedge  \rd\overline{Z}^k\wedge \cdots \wedge \rd Z^M\,.
\end{gather}
Furthermore, note that:
\begin{gather}
	\frac{\rd}{\rd t}(dp_t^j)\bigg\vert_{t=0} = \rd\left(\frac{\partial p^j}{\partial Z^k}\xi^k + F^j\right) = 0\,,
\end{gather}
where we have used~\eqref{eq:infinitesimalDiffeomorphism}. This implies that: $\rd p_t^1 \wedge \dots \wedge \rd p_t^K = \rd p^1 \wedge \dots \wedge \rd p^K$. Thus, the derivative~\eqref{eq:domegaDecomposition} satisfies the following relation:
\begin{gather}\label{eq:adjunctionOmegaExpanded}
	\frac{\rd\Omega_t}{\rd t}\bigg\vert_{t=0}	\wedge \rd p^1 \wedge \cdots \wedge \rd p^K\bigg\vert_{X} = \mathrm{Tr}~\mathrm{Jac}(\xi)~\rd Z^1 \wedge \cdots \wedge \rd Z^K\bigg\vert_{X} + \\\notag  + \sum_{j,k=1}^M\frac{\partial \xi^j}{\partial \overline{Z}^k}\rd Z^1\wedge \cdots \wedge \widehat{\rd Z^j}\wedge  \rd\overline{Z}^k\wedge \cdots \wedge \rd Z^M\bigg\vert_{X}\,.
\end{gather}
Let $\alpha \in \Gamma(X, \Omega^{n,0})$ and $\beta\in \Gamma(X, \Omega^{n-1,1})$ be the $(n,0)$ and $(n-1, 1)$ terms in the decomposition~\eqref{eq:domegaDecomposition}, respectively. Then, the forms can be expanded as:
\begin{gather}\label{eq:alphaBeta}
\alpha = f(Z)\rd Z^1\wedge \cdots \wedge \rd Z^n, \quad 
\beta = \sum_{j,k=1}^n g_k^j(Z) \rd Z^1\wedge \cdots \wedge\widehat{\rd Z^j}\wedge \cdots \wedge \rd Z^n \wedge \rd\overline{Z}^k\,.
\end{gather}
Combining~\eqref{eq:alphaBeta} and~\eqref{eq:adjunctionFormula} and solving for $f(Z)$ we obtain:
\begin{gather}\label{eq:fFunc}
	f(Z) = \mathrm{Tr}~\mathrm{Jac}(\xi)\Big/
 \det\Big({\frac{\partial p^j}{\partial Z^{n+k}}}\Big)_{j,k=1}^K\,.
\end{gather}
Similarly, solving for $g_k^j(Z)$, we obtain:
\begin{gather}\label{eq:gFunc}
	g_k^j(Z) 
 = \frac{\displaystyle (-1)^{n-j}}
        {\displaystyle \det\Big({\frac{\partial p^{j'}}
                                {\partial Z^{n+{k'}}}}\Big)_{j',k'=1}^K}
 \left[ \frac{\partial \xi^j}{\partial \overline{Z}^k} 
        +\sum_{l=1}^K\frac{\partial \xi^j}{\partial\overline{Z}^{n+l}}
                 \frac{\partial\overline{Z}^{n+l}}{\partial\overline{Z}^k}
 \right]\,,
\end{gather}
where we have applied pullback $i\colon X\hookrightarrow A$ as:
\begin{gather}
	i^*\rd\overline{Z}^{n+l}	 = \sum_{k=1}^n\frac{\partial\overline{Z}^{n+l}}{\partial\overline{Z}^k}\rd\overline{Z}^k\,.
\end{gather}
Combining the results, we see that~\eqref{eq:fFunc} and~\eqref{eq:gFunc} are natural generalizations of the results of~\cite{keller2009numerical} in the case of $K>1$.

\section{Numerical results}\label{sec:numerical}

Here we compute the Weil--Petersson metric using the Kodaira--Spencer map, and thereby obtain the physical Yukawa couplings, for a range of different complete intersection Calabi--Yau manifolds. This requires the numerical evaluation of integrals over the Calabi--Yau fibres $X_t$, which are approximated by Monte Carlo integration over $X_t$,
\begin{equation}
    \int_{X_t} \vol{g_t} 
    f \simeq \frac{1}{N} \sum_{k=1}^N f(p^{(t)}_k)\,,
\end{equation}
where the distribution of the random points $\{p^{(t)}_k\}_{k=1}^N$ is chosen to be uniform with respect to the Fubini--Study metric on the embedding space $\mathbb{P}^{n_1} \times \cdots \times \mathbb{P}^{n_K} \supset X_t$~\cite{shiffman1999distribution,braun2008calabi}. Here $N=250,000$ in all experiments unless stated otherwise. All computations are performed using our \textsf{JAX} library~\cite{Berglund:2022gvm} and some make use of the point sampling package of \textsf{cymetric}~\cite{Larfors:2022nep}. 

\subsection{Mirror of $\mathbb{P}^5[3,3]$}\label{sec:mirrorP533}
\begin{figure*}[htb]
    \centering
    \begin{subfigure}[b]{0.5\textwidth}
        \centering
        \includegraphics[width=0.95\textwidth]{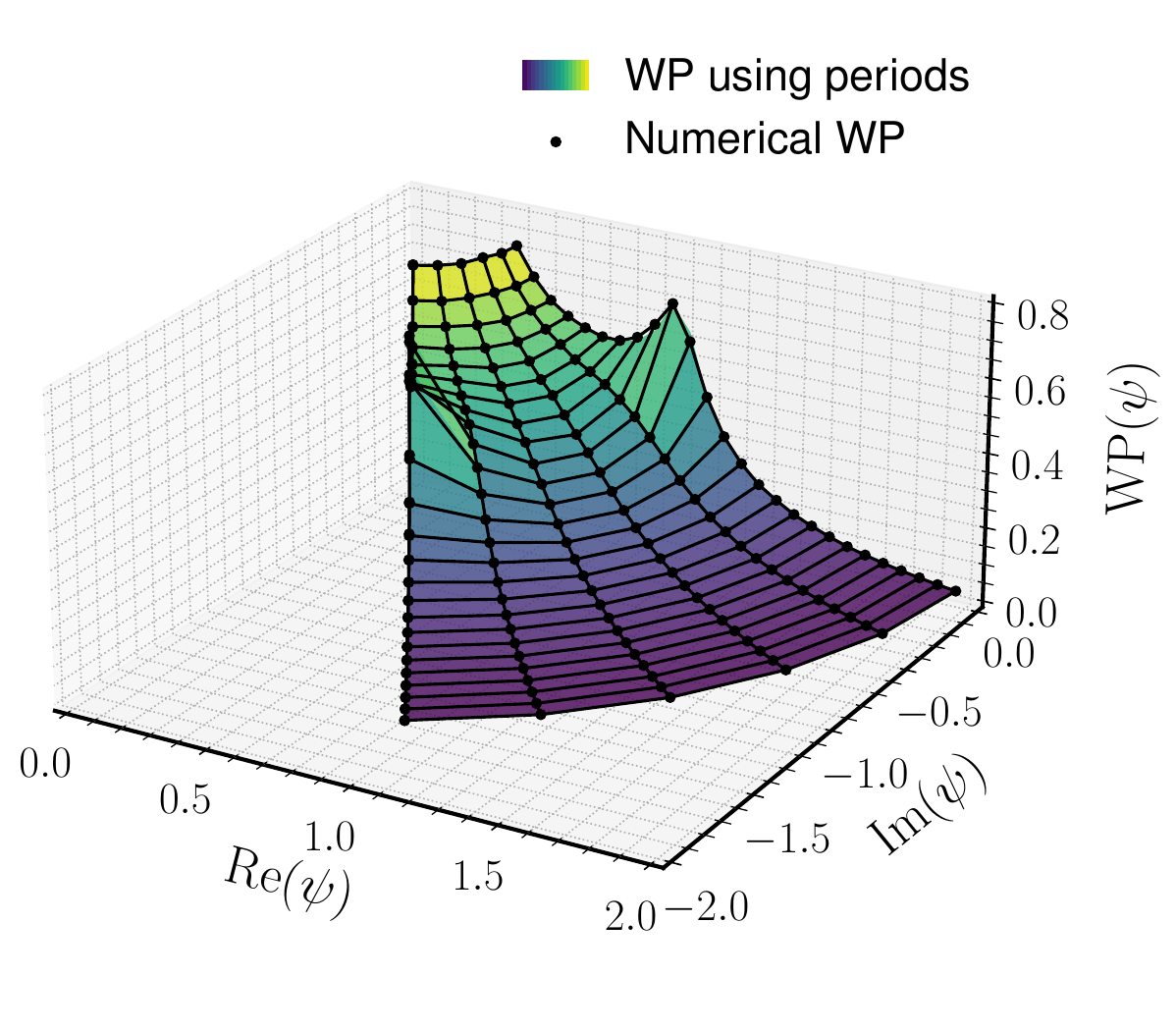}
        \capt{Weil--Petersson metric}
    \end{subfigure}%
    \begin{subfigure}[b]{0.5\textwidth}
        \centering
        \includegraphics[width=0.95\textwidth]{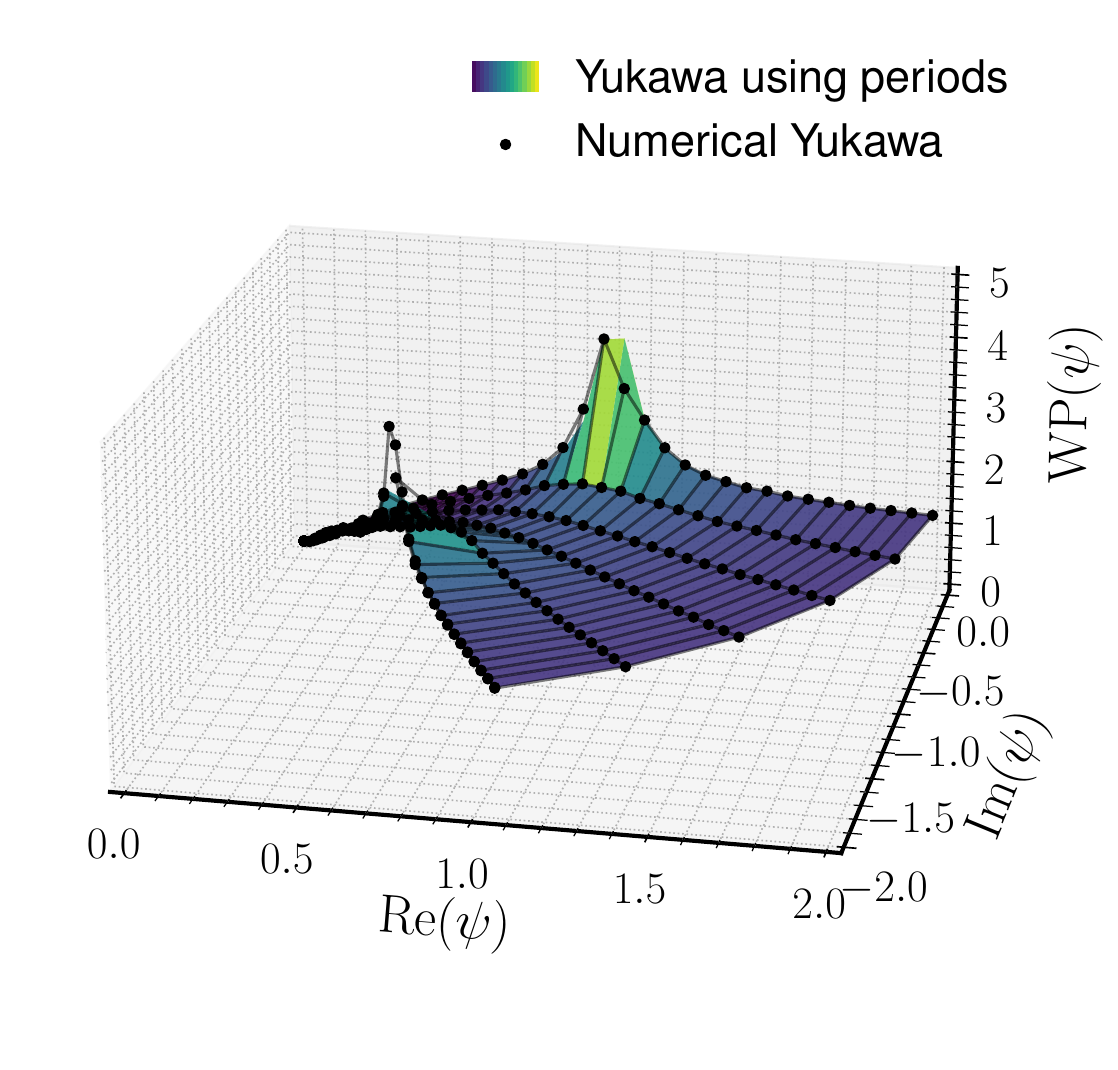}
        \capt{Normalized Yukawa coupling}
    \end{subfigure}
    \capt{Comparison of the Weil--Petersson metrics (a) and the normalized Yukawa couplings (b) on a uniform grid in $(r,\theta)$ on the disc $\{z \in \mathbb{C} \, \vert \, \vert z\vert \leq 2\}$, computed using periods with the method described in Section~\ref{ssec:constrKSmap} for the mirror of $\mathbb{P}^5[3,3]$.}
    \label{fig:X33comparison}
\end{figure*}

\begin{figure*}[htb]
 \centering
 \includegraphics[width=1.0\textwidth]{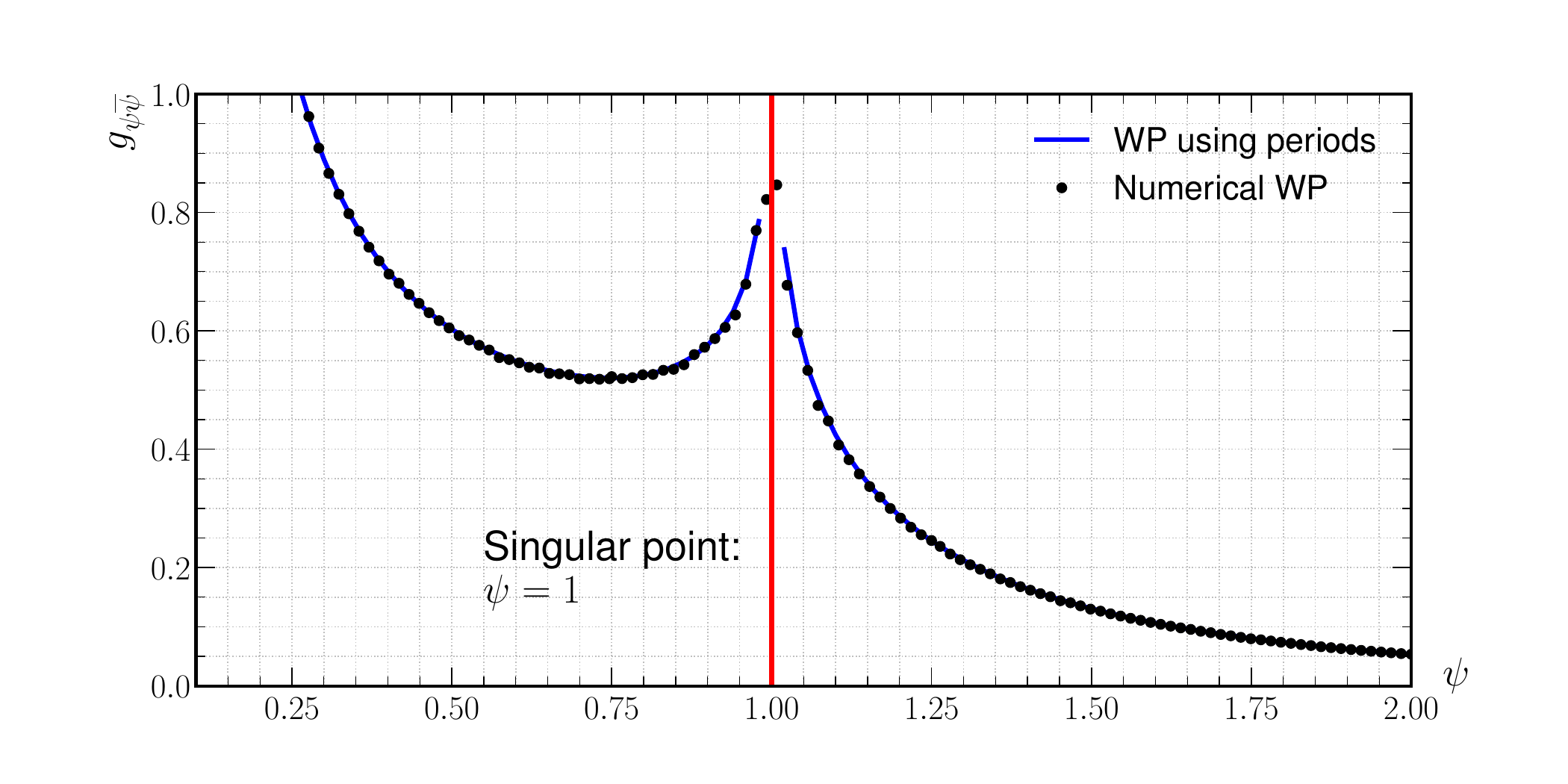}
     \capt{Comparison of the period (solid blue line) vs. the Kodaira--Spencer computation for the Weil--Petersson metric on the intersection of two cubics~\eqref{P5[33]} using points between $(0.01, 2)$ along the line ${\rm Im }(\psi)=0$. The red vertical line indicates the logarithmic singularity at $\psi=1$.}\label{fig:2DtwoCubics}
\end{figure*}

Consider the Calabi--Yau threefold $X$ belonging to the deformation space $\mathbb{P}^5[3,3]$ via the following system of defining equations:
\begin{equation}\label{P5[33]}
x_0^3 + x_1^3 + x_2^3 - 3\psi\, x_3x_4x_5 = 0 \,, \qquad
x_3^3 + x_4^3 + x_5^3 - 3\psi\, x_0x_1x_2 = 0 \,.
\end{equation}
This space has $h^{2,1}=73$ and $h^{1,1}=1$ and it is a member of the one parameter family of Calabi--Yau manifolds considered, \textit{e.g.}, in~\cite{Joshi:2019nzi}.
Its mirror, $\widetilde{X}$ (with swapped Hodge numbers), is constructed as a blowup of a finite quotient of this same zero-locus. That way, the complex structure parameter $\psi$ in the description of $X$ one-to-one corresponds to the K\"ahler class of the mirror, $\widetilde{X}$. Note however that the limit $\psi\to0$ makes the intersection of the two quadrics~\eqref{P5[33]} (as well as their finite quotient) singular along a network of curves. This singularization of both $X$ and $\widetilde{X}$ gives rise to the pole-singularity at $\psi\to0$ seen in the plots in Figure~\ref{fig:X33comparison} and Figure~\ref{fig:2DtwoCubics}. 

The grid of points in Figure~\ref{fig:X33comparison} corresponds to the Kodaira--Spencer computation, and we observe good agreement between the two methods. The modulus dependent Yukawa coupling is presented in Figure~\ref{fig:X33comparison}(b) where we again obtain a matching of the results from both techniques.  

\subsection{Quintic and the Gepner model $Y_{4;5}$}\label{sec:quintic}
One of the simplest exactly soluble models is given by a Gepner model $Y_{4;5}$~\cite{GEPNER1988757, DISTLER1988295, GEPNER1987380, Kalara:1987CP}, which corresponds to a specific point in the moduli of the quintic threefold $\mathbb{P}^4[5]$. It has been shown in~\cite{DISTLER1988295} that the normalized Yukawa couplings can be expressed as powers of a constant $\kappa$ given by a ratio of $\Gamma$-functions:
\begin{gather}
	\kappa = \left[\frac{\Gamma(3/5)^3 \Gamma(1/5)}{\Gamma(2/5)^3 \Gamma(4/5)}\right]^{1/2} \approx 1.09236 \,.
\end{gather}
In particular, in the model $Y_{4;5}$, we consider a quintic threefold $X$ defined as a zero locus of Fermat quintic $Z_0^5 + \dots + Z_4^5 = 0$ under quotient $G = \mathbb{Z}_5\times \mathbb{Z}_5'$. The resulting Calabi--Yau manifold has $h^{1,1}(X/G) = 1$ and $h^{2,1}(X/G) = 5$ and Euler number $-200/(5\times 5)=-8$ (see:~\cite{constantin2017hodge, candelas2018highly}) hence a heterotic string compactification with standard embedding will yield four chiral generations. One can show that the group $G$ is freely acting, with its generators being
\begin{align}
    \mathbb{Z}_5\,:& \quad z_j\rightarrow \alpha^j z_j\,, \\ 
    \mathbb{Z}_5^\prime\,:& \quad z_j\rightarrow z_{j+1}\,, 
\end{align}
The monomial representatives of the $H^{2,1}(X/G)\simeq H^1(T_{X/G})$ are shown in Table~\ref{tab:monomialsQuinticQuotient}.

\begin{table}[ht]\centering
\begin{tabular}{||ccc||}
\hline
      Family & Monomial $F$ & Comment \\
      \hline\hline
      $1$ & $Z_i^2 Z_j^3$  & $i \neq j$			\\
      $2$ & $Z_i Z_j Z_k^3$ & $i\neq j \neq k$  \\
      $3$ & $Z_i Z_j^2 Z_k^2$ & $i\neq j \neq k$ \\
      $4$ & $Z_i Z_j Z_k Z_l^2$ & $i \neq j \neq k \neq l$ \\
      $5$ & $Z_0 Z_1 Z_2 Z_3 Z_4$ & -- \\
      \hline
  \end{tabular}
    \capt{ Monomial representatives of $H^1(T_{X/G})$ under identification~\eqref{eq:deformationsAndCohomology} for the Gepner model $Y_{4;5}$.}\label{tab:monomialsQuinticQuotient} 
\end{table}

\begin{figure*}[ht]
    \includegraphics[width=0.4\textwidth]{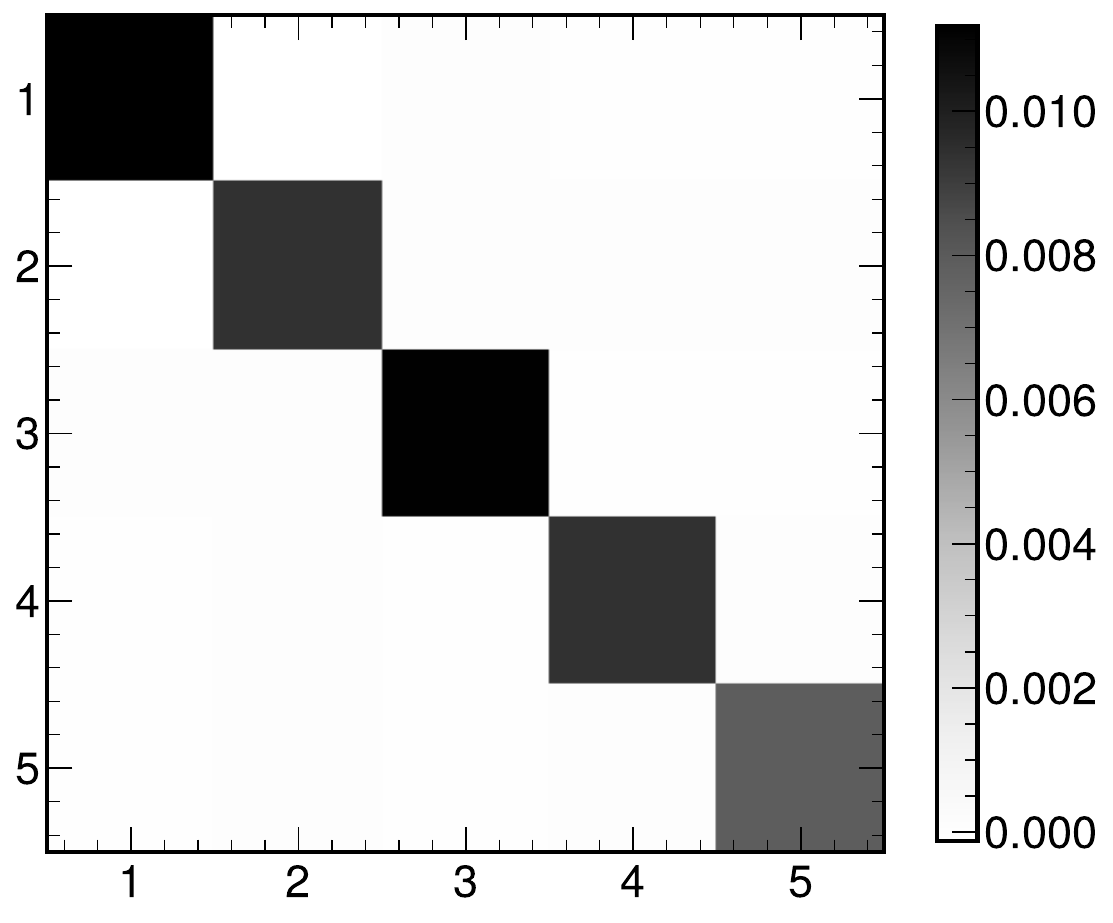}
    \centering
    \capt{Numerical values of the normalization matrix $N_{ij} = \langle b_i, b_j\rangle_\mathrm{WP}$ for the Gepner model $Y_{4;5}$. }\label{fig:normalizationMatrixQuintic}
\end{figure*}

The Yukawa couplings for the Gepner model $Y_{4;5}$ were already computed in%
~\cite{DISTLER1988295}. We want to employ the method described in Section~\ref{ssec:constrKSmap} in order to make a direct comparison. We first show that monomial families in Table~\ref{tab:monomialsQuinticQuotient} indeed form an orthogonal basis. In the basis of~\ref{tab:monomialsQuinticQuotient}, the Weil--Petersson metric entries are given in the grayscale grid of Figure~\ref{fig:normalizationMatrixQuintic}.  Evidently, the off-diagonal components vanish, wherefore the Yukawa couplings can be computed using~\eqref{eq:NormYuk}.

\begin{figure*}[ht]
    \includegraphics[width=1.0\textwidth]{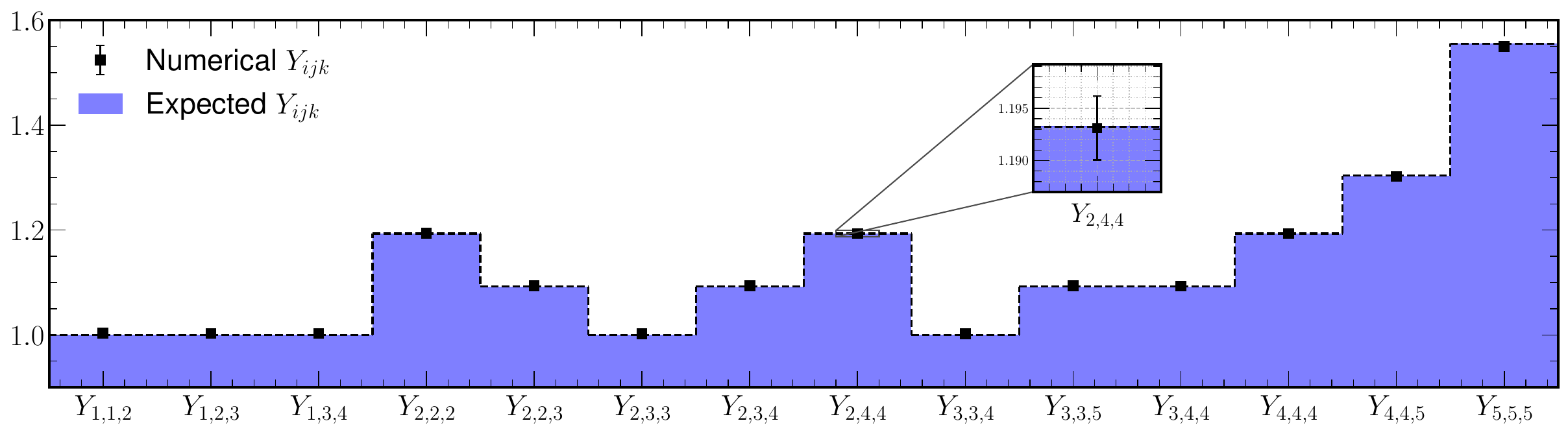}
    \centering
    \capt{Comparison of the numerical normalized Yukawa couplings of for the Gepner model $Y_{4;5}$ with the exact results in~\cite{DISTLER1988295}. The error bars indicate the $95\%$ confidence interval.}\label{fig:yukawasQuinticQuotient}
\end{figure*}

Using numerical integration techniques, we compute the $95\%$ confidence intervals of the values for the normalized Yukawa couplings for the quintic quotient model. In Figure~\ref{fig:yukawasQuinticQuotient} we contrast our results with those of~\cite{DISTLER1988295}.

As can be observed in Figure~\ref{fig:yukawasQuinticQuotient}, the numerical values computed using the methods described in the Section~\ref{ssec:constrKSmap} are within the margin of the error of the exact results computed in the work~\cite{DISTLER1988295}.

Finally, note that the coupling corresponding to the family $5$ in Table~\ref{tab:monomialsQuinticQuotient} matches the coupling of the mirror quintic $\tilde{X}$ which has $h^{2,1}(\tilde{X}) = 1$. In particular, recall that the invariant/normalized Yukawa coupling defined in~\cite{CANDELAS199121} attains value of $\kappa^5$ at Landau--Ginzburg point in the complex structure moduli of $\tilde{X}$. From Figure~\ref{fig:yukawasQuinticQuotient}, the numerical value corresponding to this family is:
\begin{gather}
	Y_{5,5,5} = 1.550\pm 0.002\,,
\end{gather}
which is close to the exact value of $\kappa^5 \approx 1.555$.

\subsection{Tian--Yau quotient}
We start with a complete intersection in $\mathbb{P}^3 \times \mathbb{P}^3$ given by the following configuration matrix 
\begin{gather}\label{eq:cmTY}
	X= \left[\begin{matrix}
		\begin{matrix}
			3 \\
			3
		\end{matrix}
		~\vline~
		\begin{matrix}
			3 & 1 & 0 \\
			0 & 1 & 3 \\
		\end{matrix}
	\end{matrix}\right]\,.%
\end{gather}
All manifolds in this deformation class have Hodge numbers $h^{1,1} = 14$ and $h^{1,2} = 23$, and so $\chi=-18$.
To be specific, we choose the defining polynomials to be of the form 
\begin{subequations}\label{CK1988}
\begin{align}
    p^1& =\tfrac13 \left(x_0^3+x_1^3+x_2^3+x_3^3\right) =0\,,\\
    p^2 & = x_0 y_0+x_1 y_1+x_2 y_2+x_3 y_3=0\,, \label{eq:p2}\\
     p^3& =\tfrac13 \left(y_0^3+y_1^3+y_2^3+y_3^3\right) =0\,,
\end{align}
\end{subequations}
where, following the notation of~\cite{CANDELAS1988357} we take $x$ and $y$ to denote coordinates in the first and second $\mathbb{P}^3$s respectively. The manifold~\eqref{CK1988} has a freely acting $\mathbb{Z}_3$ symmetry specified as follows
\begin{equation}\label{CKZ3}
\bigg\{
\begin{array}{@{}rl}
    (x_0,x_1,x_2,x_3) &\rightarrow (x_0, \alpha^2 x_1, \alpha x_2, \alpha x_3)\,,\\
     (y_0,y_1,y_2,y_3) &\rightarrow (y_0, \alpha y_1, \alpha^2 y_2, \alpha^2 y_3)\,,
\end{array}
\end{equation}
with $\alpha=e^{2\pi {\rm i}/3}$. The Tian--Yau manifold is constructed by quotienting out the freely acting $\mathbb{Z}_3$ symmetry~\eqref{CKZ3}, yielding a quotient Calabi--Yau manifold with $\chi=-6$~\cite{tian1987three,candelas2008triadophilia}.

Similarly as in the case of the quintic threefold discussed in Section~\ref{sec:quintic}, we consider the orthogonal basis of $H^1(T_{X/\mathbb{Z}_3})$ specified by the corresponding monomial representatives shown in Table~\ref{tab:monomialsTY}, constructed by Gram--Schmidt orthogonalization using the inner product defined by the 
Weil--Petersson metric.

\begin{figure*}[ht]
 \centering
 \begin{tikzpicture}
     \path[use as bounding box](-.2,0)--++(15,4.5);
     \path(7.5,2)node{\includegraphics[width=1.0\textwidth]{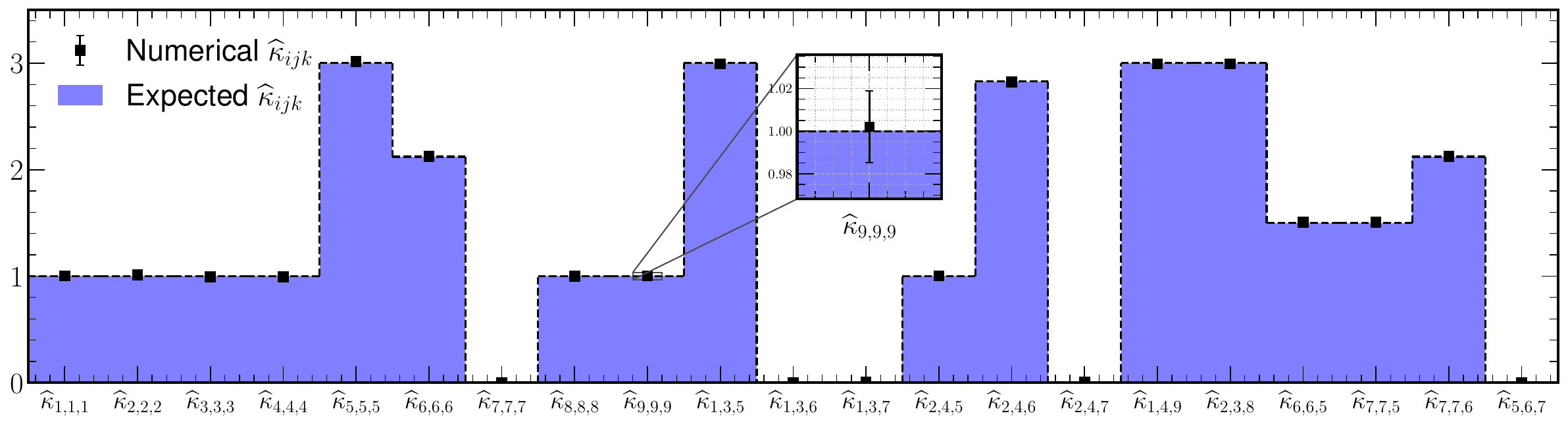}};
 \end{tikzpicture}
    \capt{Comparison of the unnormalized Yukawa couplings for the $\mathbb{Z}_3$ quotient of the Tian--Yau manifold Tian--Yau manifold with the results of~\cite{CANDELAS1988357}.  The error bars indicate the $95\%$ confidence interval.}\label{fig:yukawasTY}
\end{figure*}

\begin{table}[ht]\centering
\begin{tabular}{||cc||}
\hline
      Family & Monomial $F$  \\
      \hline\hline
      $\lambda_1$ & $x_0 x_1 x_2 \,e_1$  \\
      $\lambda_2$ & $x_0 x_1 x_3 \,e_1$ \\
      $\lambda_3$ & $y_0 y_1 y_2 \,e_3$  \\
      $\lambda_4$ & $y_0 y_1 y_3 \,e_3 $  \\
      $\lambda_5$ & $x_3 y_3 \,e_2$  \\
      $\lambda_6$ & $\displaystyle\frac{1}{\sqrt{8}}(x_3 y_3 + 3x_2 y_2)\,e_2$  \\
      $\lambda_7$ & $\displaystyle \sqrt{\frac{3}{8}}(x_0 y_0 - x_1 y_1)\,e_2$  \\
      $\lambda_8$ & $x_2 y_3 \,e_2$  \\
      $\lambda_9$ & $x_3 y_2 \,e_2$ \\
      \hline
  \end{tabular}
    \capt{Monomial representatives of $H^1(T_{X/\mathbb{Z}_3})$ considered in~\cite{CANDELAS1988357} under identification~\eqref{eq:deformationsAndCohomology}. The eigenmodes $\lambda_6$ and $\lambda_7$ are normalized with respect to the Weil--Petersson metric for $\mathbb{Z}_3$ quotient of the Tian--Yau manifold.}\label{tab:monomialsTY}
\end{table}
In Table~\ref{tab:monomialsTY} we have chosen $\{e_j\}_j$ to be such that $\mathcal{N}_{X|\mathbb{P}^3\times \mathbb{P}^3} 
 \simeq \mathcal{O}(3,0)\,e_1\oplus \mathcal{O}(1,1)\,e_2
        \oplus\mathcal{O}(0,3)\,e_3$. 
Using the numerical method discussed in Section~\ref{ssec:constrKSmap}, we compute the  unnormalized Yukawa couplings:
\begin{gather}\label{eq:kappaTY}
	\kappa_{ijk} = \int_{X/\mathbb{Z}_3} \Omega \wedge \Omega	(\lambda_i, \lambda_j,\lambda_k)\,,
\end{gather}
and verify the results by comparing the ratios to the Table~3 of~\cite{CANDELAS1988357}. In particular, the values of $\widehat{\kappa}_{ijk} \stackrel{\mathrm{def}}{=} \kappa_{ijk}/\kappa_{111}$ computed numerically using~\eqref{eq:kappaTY} and compared to~\cite{CANDELAS1988357} are shown in Figure~\ref{fig:yukawasTY}, where we observe an exact match of the results.\footnote{Except for $\kappa_{246}$, which is incorrectly given in~\cite{CANDELAS1988357} to be $0$, when it should equal $\sqrt8\,\mu$ in their notation.}   %

\begin{figure*}[ht]
    \includegraphics[width=0.4\textwidth]{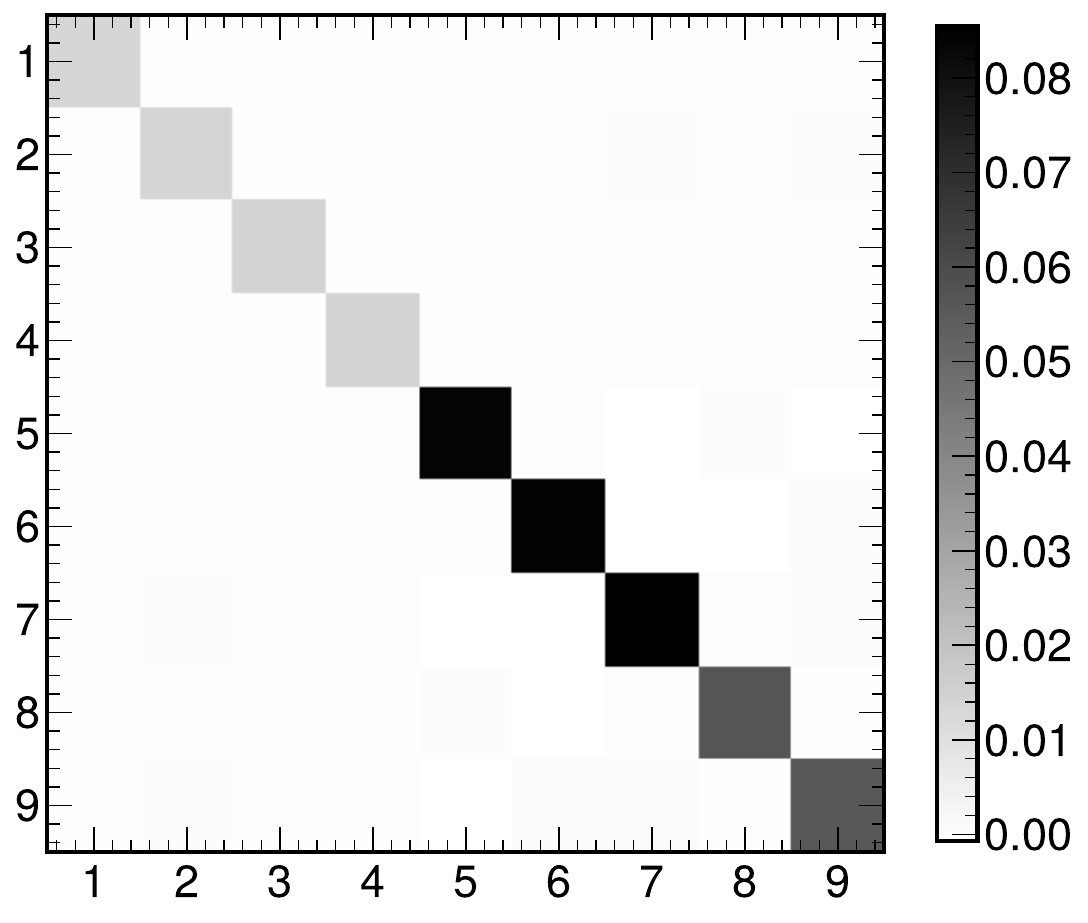}
    \centering
    \capt{Numerical values of the normalization matrix $N_{ij} = \langle \lambda_i, \lambda_j\rangle_\mathrm{WP}$  for the $\mathbb{Z}_3$ quotient of the Tian--Yau manifold~\eqref{eq:cmTY}.}\label{fig:wpTY}
\end{figure*}

\begin{figure*}[ht]
    \centering
 \begin{tikzpicture}
     \path[use as bounding box](-.2,-.4)--++(16,4.5);
     \path(7.47,1.58)node{\includegraphics[width=1.0\textwidth]{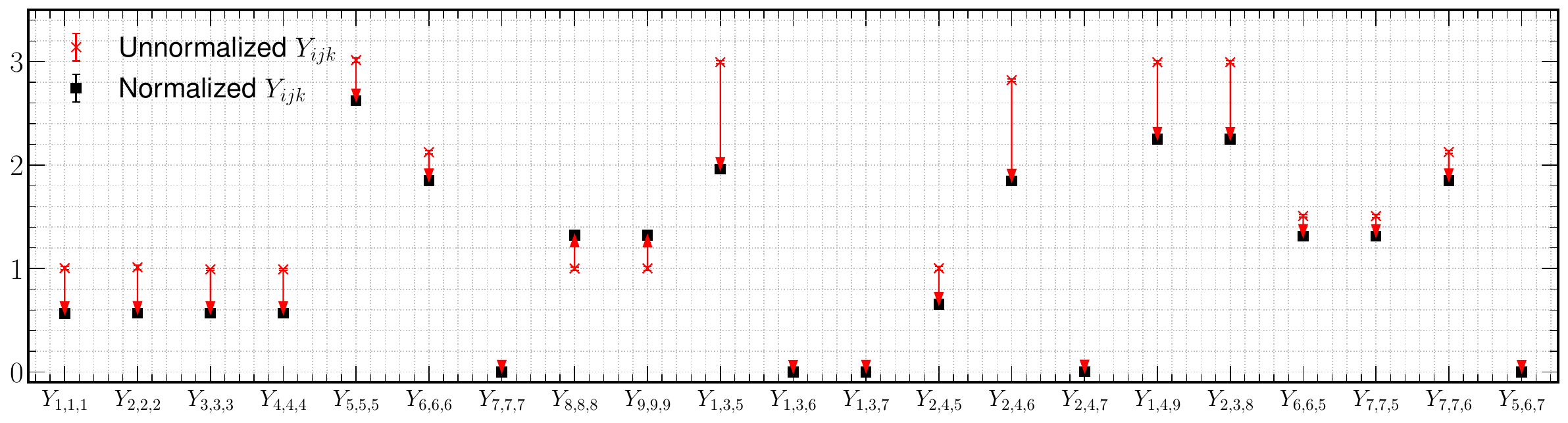}};
 \end{tikzpicture}
    \capt{The changes induced by the numerical normalization on the Yukawa couplings for the $\mathbb{Z}_3$ quotient of the Tian--Yau manifold~\eqref{eq:cmTY} are indicated by the red arrows.}
    \label{fig:normalizingYukawasTY}
\end{figure*}

Before computing the normalized Yukawa couplings, we first verify that the basis specified in Table~\ref{tab:monomialsTY} is indeed orthogonal. In particular, we plot the numerical values of the Weil--Petersson metric $N_{ij}:= \langle \lambda_i,\lambda_j\rangle_\mathrm{WP}$ in Figure~\ref{fig:wpTY}. As can be observed in Figure~\ref{fig:wpTY}, the basis $\{\lambda_j\}_j$ of $H^1(T_{X/\mathbb{Z}_3})$ is indeed orthogonal. This then allows us to use~\eqref{eq:NormYuk} to compute normalized Yukawa couplings. We show the numerical results in Figure~\ref{fig:normalizingYukawasTY}.

\begin{figure*}[ht]
 \centering
 \begin{tikzpicture}
     \path[use as bounding box](-.2,0)--++(15,6.75);
     \path(7.25,3.25)node{%
      \includegraphics%
       [viewport=0 35 900 400, clip, width=1.0\textwidth]
        {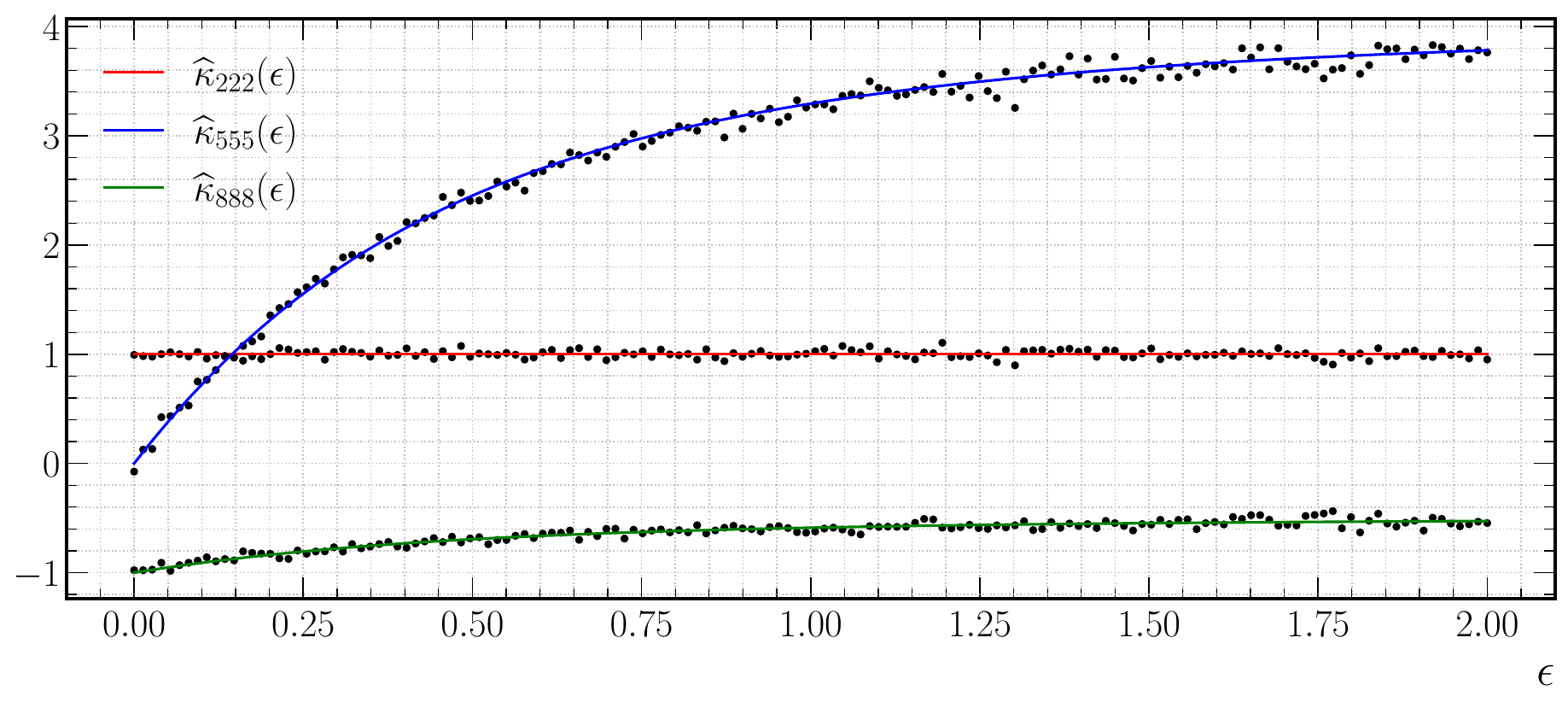}};
     \path(14.1,.85)node{\Large$\epsilon$};
 \end{tikzpicture}
    \capt{Nonzero `cubic' Yukawa couplings,
    $\widehat{\kappa}_{aaa} = \kappa_{aaa}/\kappa_{111}$ for the $\mathbb{Z}_3$ quotient of the Tian--Yau manifold using $128$ points in moduli space uniformly distributed between $(0.01, 2)$ along the line $0\leq\epsilon\leq2$. Solid lines denote expected results from~\cite{Kalara:1987qv}, while dotted points represent our numerical computation.}
    \label{fig:normalizedYukawasTY}
\end{figure*}

\begin{figure}[ht]
 \centering
 \begin{tikzpicture}
     \path[use as bounding box](-.2,0)--++(15,6.75);
     \path(7.25,3.25)node{%
      \includegraphics%
       [viewport=0 35 900 400, clip, width=1.0\textwidth]
        {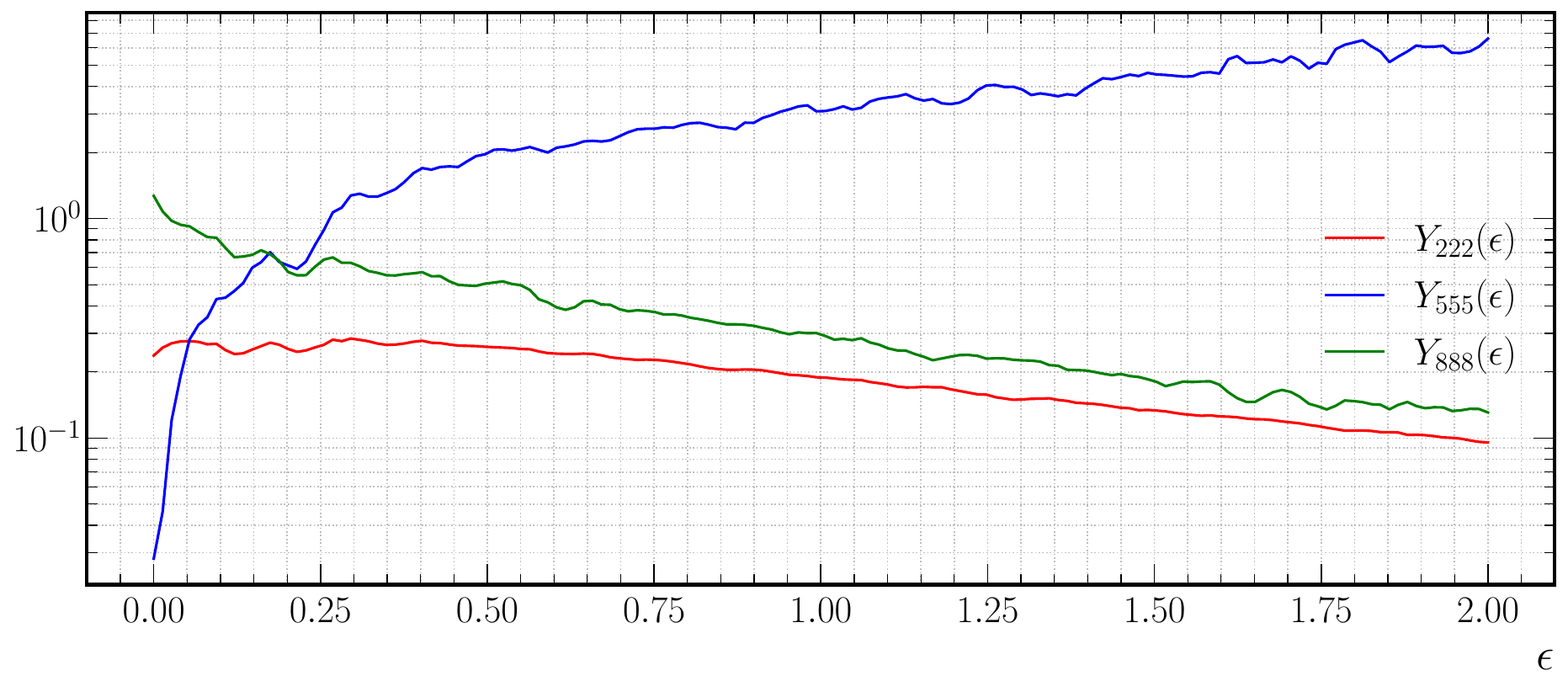}};
     \path(14.4,.85)node{\Large$\epsilon$};
 \end{tikzpicture}
   \capt{Normalized Yukawa couplings for the $\mathbb{Z}_3$ quotient of the Tian--Yau manifold, smoothed using the Savitzky--Golay filter of window size $8$ and polynomial order $3$ to remove noise due to numerical integration in the computation of the Weil--Petersson metric. Notice the onset of a large-$\epsilon$ hierarchy.}
\label{fig:normalized_YukawasTY_log}
\end{figure}

Further we compute the moduli dependent normalized and unnormalized Yukawa couplings. Consider the following deformation of $p_2$ 
(\textit{cf.},~\eqref{eq:p2} above, following~\cite[Eq.~(1)]{Kalara:1987qv}):
\begin{equation}
p_2=  x_0 y_0+x_1 y_1+(1+\epsilon)(x_2 y_2+x_3 y_3)=0\,, \quad \epsilon \in \mathbb{R}\,.
\end{equation}
In Figure~\ref{fig:normalizedYukawasTY}, we present the $\epsilon$-dependent couplings $\kappa_{aaa}$, using the basis of~\cite{Kalara:1987qv} and focusing on $0\leq\epsilon\leq2$; note that this basis differs from the one shown in Table~\ref{tab:monomialsTY}. The couplings $(\kappa_{111}, \kappa_{222}, \kappa_{333}, \kappa_{444})$ and $(\kappa_{777}, \kappa_{888})$ are respectively degenerate while the others vanish, in agreement with the results in~\cite{Kalara:1987qv}. In Figure~\ref{fig:normalized_YukawasTY_log}, we plot the absolute value of the normalized Yukawa couplings on a logarithmic scale.\footnote{Plotting absolute values shows an artificial ``crossover'' near $\epsilon\approx0.20$: In fact, $Y_{888}$ and $Y_{555}$ have opposite signs and never coincide. In turn, the physical normalization moves the actual $Y_{222}/Y_{555}$ ``crossover'' coincidence from $\epsilon\approx0.15$ in Figure~\ref{fig:normalizedYukawasTY} to $\epsilon\approx0.05$ in Figure~\ref{fig:normalized_YukawasTY_log}.}

Post-normalization, we note that as $\epsilon$ increases, a hierarchy between $Y_5=Y_{555}$ and the other couplings develops,\footnote{Similar hierarchies appearing after normalization have been observed for toroidal orbifolds~\cite{Ishiguro:2021drk}.} and they appear to converge to different values in the large-$\epsilon$ regime. Looking at the logarithmic plot of the absolute values for the normalized Yukawa couplings we identify a hierarchy of $10^2$ in the couplings. We defer a further analysis of the phenomenological implications of this large-$\epsilon$ hierarchy, as well as the $\widehat\kappa_{222}/\widehat\kappa_{555}$``crossover'' near $\epsilon\approx0.15$ in Figure~\ref{fig:normalizedYukawasTY} to a later effort; for an early phenomenological discussion at the level of the unnormalized Yukawa textures, see also~\cite{Kalara:1987qv}. 

\section{Machine learning harmonic representatives}\label{sec:mlh}

\subsection{Preliminaries}

Given a Calabi--Yau $(X,g,\omega)$, we wish to obtain an expression for forms which are harmonic with respect to the unique Ricci-flat metric $g_t$ in each Kähler class. Consider the complex analytic family $(\mathcal{X}, B, \pi)$ such that $\pi^{-1}(0) \define X_0 = X$. Any two fibres in this family are diffeomorphic as real manifolds, $X_0 \simeq X_t$. The corresponding generator $\xi$, of the infinitesimal diffeomorphism is a non-holomorphic section of $T_{X_0}$. In what follows we let $(E,G)$ denote a holomorphic vector bundle over $X$ equipped with a Hermitian structure, which may be considered the data of a smoothly varying Hermitian inner product $G$ on each fibre $E_x$. Recall $\bar{\partial}_E: \Omega^{p,q}(E) \rightarrow \Omega^{p,q+1}(E)$ is the natural generalisation of the Dolbeault operator to holomorphic bundle-valued forms $\Omega^{p,q}(E)$. For the case where $E$ is the holomorphic tangent bundle $T_X$, the Kähler metric $g$ plays the role of the fibre metric $G$. In what follows we identify $E$ with $T_X$, although in principle our method may be reproduced for an arbitrary holomorphic vector bundle if the Hermitian inner product $G$ is known. 

We may obtain a basis for $T_t B \simeq H^1(X, T_X)$ by first computing $\xi(z,\bar{z})$ in local holomorphic coordinates $\{z^i\}_{i=1}^n$ on $U \subset X_0$. Then one obtains a representative of the corresponding Kodaira--Spencer class as
$\phi = \phi^{\mu}_{\overline{\nu}} \, \rd\bar{z}^{\overline{\nu}} \otimes \pdv{}{z^{\mu}} \define \bar{\partial}_{E} \xi \in H^1(X, T_X)$, where $\phi^{\mu}_{\overline{\nu}}\,\rd \bar{z}^{\overline{\nu}} = \bar{\partial} \xi^{\mu}$. Note $[\phi] \neq 0$ as $\xi$ is not globally defined. By Hodge theory, every cohomology class contains a unique harmonic representative $\eta$, related to $\phi$ through an $\bar{\partial}_{E}$-exact correction:
\begin{equation}
 \eta = \phi + \bar{\partial}_{E}\,\mathfrak{s}
        \in \mathcal{H}_{\bar{\partial}}^{0,1}(X, T_X)\,, \quad 
        \mathfrak{s} \in C^{\infty}(T_X)\,,
\end{equation}
where $\mathfrak{s}$ is a non-holomorphic section of $T_X$. Note that $\bar{\partial}_E\,\eta = 0$ locally by construction and thus the harmonicity condition reduces to $\bar{\partial}_E^{\dagger} \eta =0$, which we encode numerically. 

\subsection{Harmonic objective}

To recover the $(0,1)$ $T_X$-valued forms which are harmonic with respect to $g_t$, there are a range of possibilities to pursue, owing to the rich interplay between geometry, topology and analysis realized in Hodge theory. Here $\eta$ is obtained through a straightforward two-stage process. First, we approximate the Ricci-flat metric for a given choice of complex structure on $X$. Secondly, we fix the learned metric on $X_t$ and parameterize the sections $s_{\theta}$ corresponding to the basis for $T_tB$ using a neural network with parameters $\theta$. Noting that $\bar{\partial}_E^{\dagger} \eta \in C^{\infty}(T_X)$, we use the natural objective for training:
\begin{align}\label{eq:harmonicLoss}
    \ell(\theta) &:= \left(\bar{\partial}_E^{\dagger} \eta_{\theta}, \bar{\partial}_E^{\dagger} \eta_{\theta}\right) = \int_{X_t} \bar{\partial}_E^{\dagger} \eta_{\theta} \wedge \bar{\star}_E  \bar{\partial}_E^{\dagger} \eta_{\theta}\,, \\
    &= \int_{X_t} \vol{g_t} \, g_{\mu \overline{\nu}} (\bar{\partial}_E^{\dagger} \eta_{\theta})^{\mu} (  \overline{\bar{\partial}_E^{\dagger} \eta_{\theta}}  )^{\overline{\nu}}\,.
\end{align}

Note the expression for the Weil--Petersson metric (\ref{eq:dOmegaResultHarmonic}) simplifies to the pairing between the interior product of the respective harmonic representatives with the holomorphic $(n,0)$ form if the Kodaira--Spencer representative is chosen to be harmonic. For the case of a general gauge bundle where a representative of the $H^1(X,E)$ cohomology may not be available, one would have to enforce the closure condition $\bar{\partial}_E \eta = 0$ in addition to the co-closure condition. 

Note that the Ricci-flat metric $g_t$ does double duty in the case of the standard embedding as it naturally induces a metric on $\Omega^{0,1}(T_{X_t})$. To find harmonic bundle-valued representatives for a general holomorphic vector bundle $V$ using the objective~\eqref{eq:harmonicLoss}, one must compute the Hermitian metric on the fibres of $V$ in addition to the metric $g_t$.

\begin{figure}[ht]
 \centering
 \includegraphics[width=1.0\textwidth]{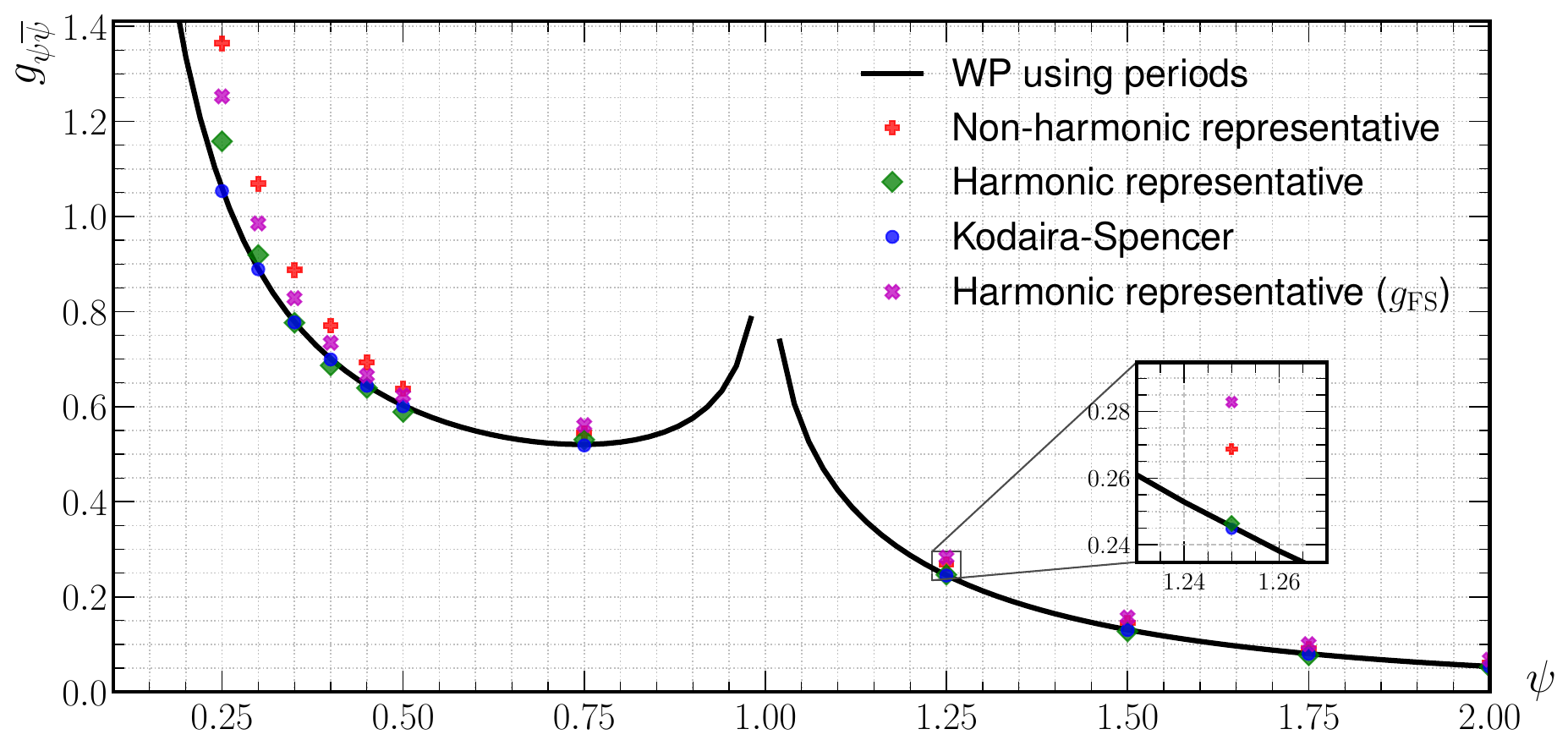}
  \capt{Weil--Petersson metric computed for the mirror of the intersection of two cubics~\eqref{P5[33]} along $\Im(\psi)=0$, obtained from the Kodaira--Spencer map, the period integral and the machine-learned harmonic representative method; \textit{cf.}\ Figure~\ref{fig:2DtwoCubics}. Red \textcolor{red}{\texttt{+}} - labeled points indicate computation of the intersection pairing~\eqref{eq:dOmegaResultHarmonic} using only the non-harmonic `reference' representatives of the Kodaira--Spencer class.}
\label{fig:eta}
\end{figure}

\begin{figure}[ht]
 \centering
 \includegraphics[width=0.9\textwidth]{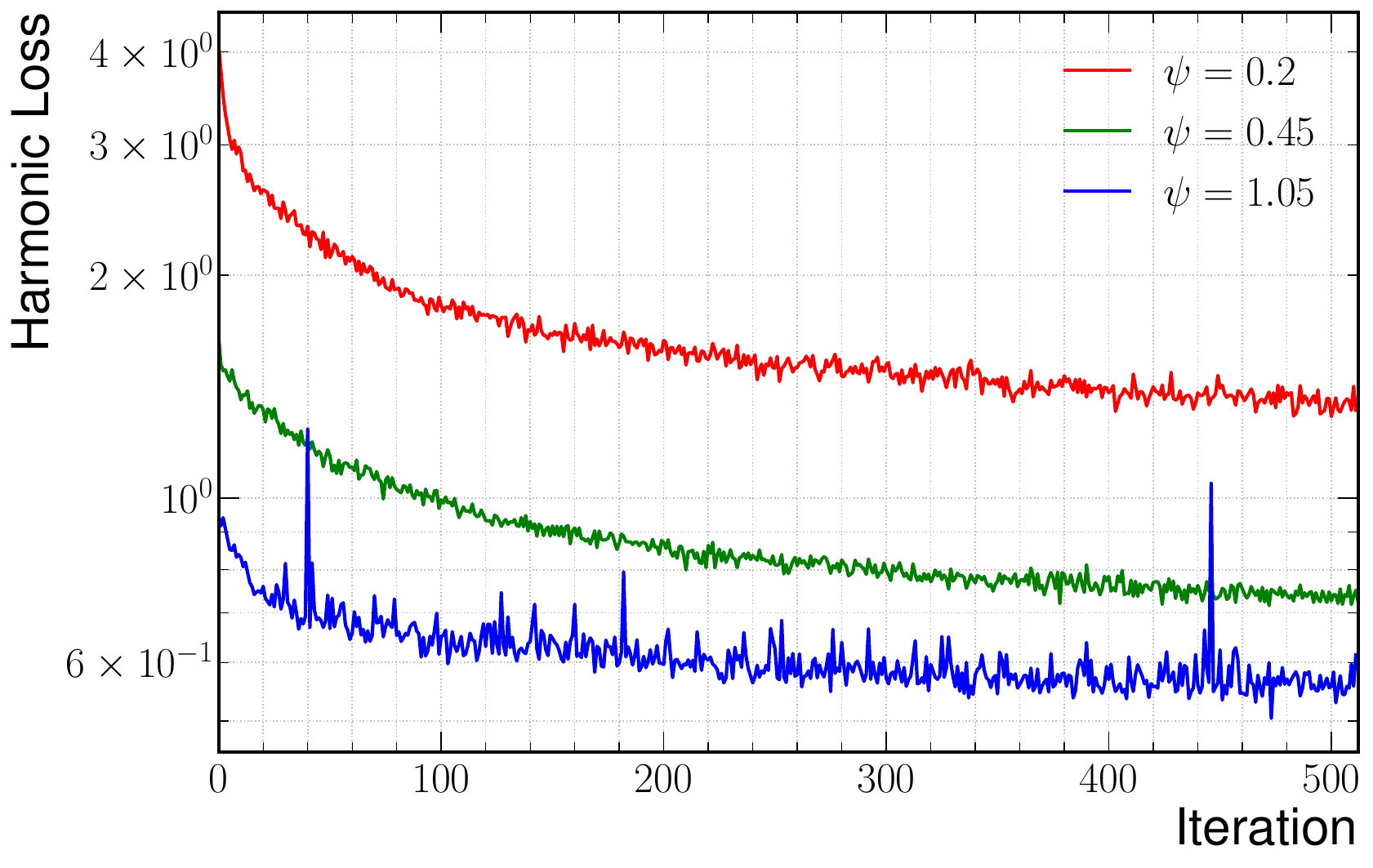}
  \capt{Evolution of the loss function for harmonic representative during the training for the mirror of $\mathbb{P}^5[3,3]$, Eqs.~\eqref{P5[33]}, at several values of $\psi$. Note that the values of the harmonic loss (the co-closed condition Eq.~\eqref{eq:harmonicLoss}) at different points in moduli space are not strictly comparable as this merely measures the vanishing condition $\Delta_{g_{\psi \overline{\psi}}} \eta = 0$ for different values of $\psi$, and the fidelity of the metric to the unique Ricci-flat metric varies as a function of $\psi$, for an identical computational budget at each point. Just because a metric better satisfies the harmonicity condition w.r.t. $g_{\psi \overline{\psi}}$ does not mean that it will yield a more accurate prediction of physical observables, as this also depends on the 'accuracy' of the metric.}\label{fig:P533HarmonicLoss}
\end{figure}

As an example, we consider the mirror $\tilde{X}$ of $X = \mathbb{P}^5[3,3]$, described in Section~\ref{sec:mirrorP533}, paramet\-ri\-zed by the complex structure parameter $\psi$. We first find explicit harmonic representatives of the Kodaira--Spencer class $[\overline{\partial}\xi]$ via the objective~\eqref{eq:harmonicLoss} using a standard fully connected neural network with four layers with intermediate dimensions $[64,32,32,48]$ and the Gaussian error linear unit activation function. We used the Adam optimization algorithm with a learning rate of $10^{-4}$ in all experiments. We empirically observed that the results were insensitive to the choice of hyperparameters considered. We subsequently compute the Weil--Petersson metric $g_{\psi \overline{\psi}}$ on an independent validation set consisting of 250,000 fibre points for each point in moduli space. In Figure~\ref{fig:eta}, we plot $g_{\psi \overline{\psi}}$ obtained using the intersection pairing valid for harmonic representatives~\eqref{eq:dOmegaResultHarmonic}, together with the numerical values obtained for general representatives by the Kodaira--Spencer approach, as well as the exact period computation across discrete points in complex moduli space. The evolution of the loss function \eqref{eq:harmonicLoss} throughout the training is shown on Figure~\ref{fig:P533HarmonicLoss} for the benchmark points $\psi=0.2\,,0.45\,,1.05$.

We observe that the harmonic forms computed using the approximate Ricci-flat metric as input are able to recover the value of the Weil--Petersson metric away from the singularities in moduli space. This supplies an additional reassurance similar in spirit to the topological computations from local geometry in~\cite{Berglund:2022qsb} ---  that numerical approximations to the Ricci-flat metric are able to yield physically meaningful data. While we have restricted our attention to the standard embedding, these results are an encouraging step towards conducting similar computations for more general gauge bundles to extract relevant effective field theory parameters. Our method provides an improved approximation to the Weil--Petersson metric over the non-harmonic representatives even near points in moduli space of the mirror $\tilde{\mathbb{P}^5[3,3]}$ close to the singularity at $\psi=0$, %
and continues to yield the expected result near logarithmic singularities such as $\psi=1$. However, the fidelity of approximation degrades as one approaches the $\psi=0$ singularity; we aim to address this issue in ongoing work.

Finally, we also consider the Weil--Petersson metric of the mirror quintic $\tilde{\mathbb{P}^4[5]}$ at Fermat point ($\psi=0$). From the machine learned harmonic representative, we obtain the value: $g_{\psi\overline{\psi}}=0.1906$, which is well within the typical $95\%$ confidence interval of the numerical integration of the exact value \cite{CANDELAS199121}:
\begin{gather}
	25\frac{\Gamma(4/5)^5\Gamma(2/5)^5}{\Gamma(1/5)^5\Gamma(3/5)^5}\approx 0.1922 \,.
\end{gather}

\section{Discussion and outlook}\label{sec:discussion}
The as yet unrealized dream of string phenomenology is to start from a construction involving higher dimensional geometry and objects such as strings and branes to obtain an effective four-dimensional theory satisfying all known particle physics and cosmological constraints from real world experiments and observations.
Plausibly, we must look for Standard Model or beyond the Standard Model physics as an $\mathcal{N}=0$ (non-supersymmetric) quantum field theory in a de~Sitter background.
With our current understanding of string theory, there is no obvious direct attack for achieving this goal in one fell swoop.
A more tractable initial step is to obtain a four-dimensional $\mathcal{N}=1$ theory that includes the Standard Model spectrum and interactions and has Yukawa couplings commensurate with observed mass hierarchies in a three generation model.
In order to perform this set of calculations within a heterotic compactification on a Calabi--Yau threefold, we require knowledge of the Ricci-flat metric in a given K\"ahler class as a function of the complex structure moduli.
This is necessary \textit{ab initio} for normalizing the kinetic terms in the K\"ahler potential and ultimately for addressing more complicated issues such as moduli stabilization, incorporating $\alpha'$-corrections, and breaking supersymmetry.
Perhaps the most straightforward laboratory for studying the problem invokes the ``standard embedding,'' wherein the holomorphic vector bundle is the tangent bundle on the Calabi--Yau manifold.
The important technical simplification that occurs within this setting is that the normalized Yukawa couplings can be computed directly using the Weil--Petersson metric on the complex structure moduli space.
In this work, we have performed the analysis for several Calabi--Yau geometries and calculated the normalized Yukawa couplings.

The Kodaira--Spencer approach~\cite{keller2009numerical} we have used applies to Calabi--Yau geometries with arbitrary $h^{2,1}(X)$.
Moreover, we can generalize these techniques to Calabi--Yau threefolds realized as hypersurfaces in toric varieties, namely those geometries obtained from applying Batyrev's procedure~\cite{Batyrev:1993oya} to the Kreuzer--Skarke list~\cite{Kreuzer:2000xy} of four-dimensional reflexive polytopes.

In particular, we are performing analogous computations of Yukawa couplings in multi-parameter families and for various three generation models such as~\cite{Greene:1986ar,Greene:1986bm,Greene:1986jb, Schimmrigk:1987ke, candelas2008triadophilia}.
The results are compared to Yukawa coupling calculations reliant on machine learned harmonic representatives using the Ricci-flat Calabi--Yau metric.
This work is forthcoming~\cite{wip}.
We as well intend to adapt these methods to non-standard embeddings, for which the number of generations of particles in the low-energy spectrum need not be given by the Euler characteristic of the Calabi--Yau base space.

Accompanying this work and our earlier paper~\cite{Berglund:2022gvm}, we aim to release our code base, the software library \textsf{cymyc}~\cite{cymyc}, written in \textsf{JAX}, to compute \underline{C}alabi--\underline{Y}au \underline{M}etrics, \underline{Y}ukawas, and \underline{C}urvature.
On complete intersection Calabi--Yau manifolds, the spectral networks we employ supply, to date, the most efficient tool for numerically approximating the Ricci-flat metric using  $\sim 10^5$ or $\sim 10^6$ points.\footnote{
The point selection follows the prescription of Shiffman--Zelditch~\cite{shiffman1999distribution}.
Preliminary experiments indicate that computational performance may be improved with different point selection schemes.}

The numerical calculation of the Weil--Petersson metric for arbitrary number of complex structure moduli will also be useful for studying the swampland distance conjecture~\cite{Ooguri:2006in}. Recently there has been some progress in this direction, by employing the Kodaira--Spencer method~\cite{keller2009numerical} for studying the moduli metric on Fermat quintic~\cite{Ashmore:2021qdf}.

A synoptic view of this research places it in the broader context of a Big Data approach to string phenomenology and the vacuum selection problem.
We envision a systematic search through the estimated mole of Standard Model-like string constructions~\cite{Constantin:2018xkj} arising from complete intersection Calabi--Yau geometries and the heptagoogol moles ($10^{700+23}$) of toric ones so as to find ``the needle in the haystack,'' which is us, living in our Universe.
To make progress, we must incorporate some combination of the algebro-geometric and analytic methods into the mechanized algorithm.
Each configuration is an entire (continuous) deformation space of models, so it seems important to obtain the physically normalized Yukawa couplings as functions of the complex structure deformation parameters and the K\"ahler class of the metric.
Ideally, we would input a Calabi--Yau geometry and ask whether there exists a point in its moduli space that recovers a quantum field theory with desired phenomenological properties and if so to deduce its low-energy effective action upon compactification.
We suspect that string Standard Models with hierarchies in the Yukawa couplings are extremely rare, as at a generic point in moduli space, most of the couplings will be of $\mathcal{O}(1)$.
Finding special points in moduli space where this expectation is dashed poses a central challenge for obtaining realistic models of particle physics.
Given the vastness of potential compactification geometries, it must be the case that if we find one model with the correct physics, there will be hugely many.

More ambitiously, the question of whether Calabi--Yau compactifications (\textit{a priori}, with Minkowski spacetime) can be \emph{uplifted} to accommodate our asymptotically de~Sitter spacetime  may well depend ``Goldilocks'' style, delicately on
\emph{nearly-but-not-quite/almost-conifold-singular} Calabi--Yau manifolds~\cite{Bento:2021nbb,Berglund:2022qsb}.
Our earlier machine learning investigations scanning for curvature clumping on singular K3 surfaces~\cite{Berglund:2022gvm} (generalized to Calabi--Yau threefolds), should provide a solid stepping stone in searching for such models.

\section*{Acknowledgements}
We thank Nana Cabo Bizet and Fernando Quevedo as well as the organizers and participants at String Data 2023 at Caltech for comments on this work.
PB and GB are supported in part by the Department of Energy grant DE-SC0020220.
 TH\ is grateful to the Department of Mathematics, University of Maryland, College Park MD, 
 and the Physics Department of the Faculty of Natural Sciences of the University of Novi Sad, Serbia, for the recurring hospitality and resources.
VJ is supported by the South African Research Chairs Initiative of the Department of Science and Innovation and the National Research Foundation. DM is supported
by FCT/Portugal through CAMGSD, IST-ID, projects UIDB/04459/2020 and UIDP/04459/2020. 
CM is supported by a Fellowship with the Accelerate Science program at the Computer Laboratory, University
of Cambridge.
JT is supported by a studentship with the Accelerate Science Program. 
The authors would like to thank the Isaac Newton Institute for Mathematical Sciences for support and hospitality during the program ``Black holes: bridges between number theory and holographic quantum information'' when work on this paper was undertaken; this work was supported by EPSRC grant number EP/R014604/1.

\appendix
\section{Proofs of lemmas}\label{sec:proofs}

For completeness, we provide proofs of Lemma~\ref{lemma:Serre} and Lemma~\ref{lemma:WPHarmonic}.
We recall that we have used Lemma~\ref{lemma:Serre} to establish that $H^1(X, T_X)\simeq H^{2,1}(X)$ and Lemma~\ref{lemma:WPHarmonic} to describe the Weil--Petersson metric.

\begingroup
\noindent
{\bf Lemma~\ref{lemma:Serre}}
~~\it
Let $X$ be a Calabi--Yau manifold, then: $H^1(X,T_X)\simeq H^{n-1,1}_{\overline{\partial}}(X)$ and the isomorphism is given by:
\begin{gather}
	[\alpha] \longmapsto [\Omega(\alpha)]\,,
 \tag{\ref{e:Serre}$'$}
\end{gather}
where $\Omega\in H^{n,0}_{\overline{\partial}}(X)$ is non-zero.
\endgroup

\pf{We shall first show the isomorphism is a consequence of Serre duality. For brevity we shall suppress the reference to the manifold $X$. Then, by multiple applications of Serre duality, we obtain:\begin{gather}
	H^1(T_X) \simeq H^{0,1}(T_X) \simeq H^{n, n-1}(T_X^*)^* \simeq H^{n-1}(\Omega^n\otimes T_X^*)^*\,.
\end{gather}
However, since $\Omega^n = K_X \simeq \mathcal{O}_X$ and $T_X^*\simeq \Omega$, we have:
\begin{gather}
	H^{n-1}(\Omega^n\otimes T_X^*)^* \simeq H^{1,n-1}(\mathcal{O}_X)^* \simeq H^{n-1,1}(\mathcal{O}_X^*)\simeq H^{n-1,1}(\mathcal{O}_X)\,,
\end{gather}
thus proving the claim $H^1(T_X) \simeq H^{n-1,1}_{\overline{\partial}}(X)$. We shall now show that the isomorphism is given by the interior product. It is trivial to see that the interior product gives an isomorphism on the sections: $\Gamma(X, \Omega^{n-1,1}) \simeq \Gamma(X, \Omega^{0,1}(T_X))$~\cite{tian:1987}, thus what remains to show is that the map preserves kernels and images of ${\overline{\partial}}$. Let $p\in X$ and pick local coordinates $(z^1,\dots, z^n)$ centered at point $p$. Suppose that $[\alpha] = 0$ in $H^1(T_X)$, then, $\alpha = \overline{\partial}\phi$ for some non-holomorphic vector field $\phi \in \Gamma(X,T_X)$. Let $\Omega = f(z)\rd z^1\wedge \cdots \wedge \rd z^n$ in the local coordinates defined above, where $f(z)$ is holomorphic. Then, the interior product is given explicitly by:
\begin{gather}
	\Omega(\overline{\partial}\phi) = \sum_{\mu=1}^n (-1)^{\mu-1} f\, \alpha^{\mu}_{\overline{\nu}}\, \rd\overline{z}^\nu \wedge \widehat{\mu}
 = \sum_{\mu=1}^{n}(-1)^{\mu-1} f\, \frac{\partial \phi}{\partial \overline{z}^\nu}\rd\overline{z}^\nu\wedge \widehat{\mu}\,,
\end{gather}
where we abbreviated 
$\widehat{\mu} \overset{\scriptscriptstyle\mathrm{def}}
=dz^1\wedge \cdots \wedge \widehat{dz^\mu}\wedge \cdots \wedge dz^n$.
That above is exact follows immediately from the fact that $f$ is holomorphic, thus $\overline{\partial}f = 0$. Similarly, let $\alpha$ be $\overline{\partial}$-closed, then:
\begin{gather}
	\overline{\partial}\Omega(\alpha) 
 = \sum_{\mu=1}^{n}(-1)^{\mu-1}	f\,\frac{\partial \alpha_{\overline{\nu}}^\mu}{\partial \overline{z}^\sigma} \rd\overline{z}^\sigma \wedge \rd\overline{z}^\nu \wedge \widehat{\mu} 
 = 0\,.
\end{gather}
}

\begingroup
\noindent
{\bf Lemma~\ref{lemma:WPHarmonic}}~~\it
Let $\Omega \in H_{\overline{\partial}}^{n,0}(X_t)$ where $n=\dim{X_t}$, be non-zero, then:
\begin{gather}
	\langle a, b\rangle_{\mathrm{WP}} = - \frac{\displaystyle \int_{X_t} \Omega(\mathcal{H}\rho(a)) \wedge \overline{\Omega(\mathcal{H}\rho(b))}}{\displaystyle \int_{X_t} \Omega \wedge \overline{\Omega}} \int_{X_t}\vol{g_t}\,.
 \tag{\ref{L:WPH}$'$}
\end{gather}
\endgroup
\pf{
The result follows from direct calculation in local coordinates. Let $p\in X_t$ and consider local coordinates $(z^1, \dots, z^n)$ centered at $p$. Let $\alpha = \mathcal{H}\rho(a)$ and $\beta = \mathcal{H}\rho(b)$, then:
\begin{gather}
	\Omega(\alpha) = \sum_{\mu,\nu=1}^n (-1)^{\mu-1}	 \alpha_{\overline{\nu}}^{\mu}\, f\, \rd\overline{z}^\nu \wedge \widehat{\mu}\,,
\end{gather}
where $\alpha = \alpha^{\mu}_{\overline{\nu}}\,\rd\overline{z}^{\nu}\otimes \partial_{\mu}$ and $\Omega = f(z) \, dz^1\wedge \cdots \wedge dz^n$. Combining the results for both deformations yields the following:
\begin{gather}
	\Omega(\alpha)\wedge\overline{\Omega(\beta)}	 = \sum_{\mu,\mu',\nu,\nu'=1}^n (-1)^{\mu+\mu'}|f|^2\,\alpha_{\overline{\nu}}^{\mu}\,\overline{\beta_{\overline{\nu'}}^{\mu'}}\, \rd \overline{z}^{\nu}\wedge \widehat{\mu} \wedge \rd z^{\nu'}\wedge \overline{\widehat{\mu'}}\,.
\end{gather}
It is easy to see that the only non-vanishing terms have indices $\mu = \nu'$ and $\mu' = \nu$, which leads to the following expression:
\begin{gather}
	\Omega(\alpha)	\wedge \overline{\Omega(\beta)} 
 = \left(\sum_{\mu,\nu=1}^{n}(-1)^{\mu+\nu}\alpha_{\overline{\nu}}^{\mu}\,
 \overline{\beta_{\overline{\mu}}^{\nu}}\,
 (-1)^{\mu+\nu-1}\right)\Omega\wedge \overline{\Omega} 
 = -\alpha_{\overline{\nu}}^{\mu}\,\overline{\beta_{\overline{\mu}}^\nu}\,
 \Omega\wedge\overline{\Omega}\,.
\end{gather}
Finally, since $\alpha$ and $\beta$ are both chosen to be harmonic, using the results of~\cite{nannicini:1986} 
(or Lemma~\ref{lemma:harmonicNannicini}), it is immediate that $\alpha_{\overline{\nu}}^{\mu}\,\overline{\beta_{\mu}^{\nu}} = g_{\mu\overline{\nu}}\, g^{\sigma \overline{\delta}}\, \alpha^{\mu}_{\overline{\delta}}\,
\overline{\beta_{\overline{\sigma}}^\nu}$ where (see Definition~\ref{defWP}) 
$g_{t} = g_{\mu\overline{\nu}}\,\rd z^\mu\wedge \rd\overline{z}^{\nu}$. Furthermore, noting that the flat metric on compact $X_t$ solves the Monge--Amp\`ere equation: $\vol{g_t} = \kappa\, \Omega\wedge \overline{\Omega}$ for some constant $\kappa\in\IC$, we obtain:
\begin{gather}
	\int_{X_t}\Omega(\alpha)\wedge\overline{\Omega(\beta)}	 = - \frac{1}{\kappa}\int_{X_t}\alpha \wedge \overline{\star_{g_t}}\beta\,.
\end{gather}
After identifying $\kappa$ with $\int_{X_t}\vol{g_t} \big/ \int_{X_t}\Omega\wedge\overline{\Omega}$, the result follows.
}

We also note that there exists a method of computation of the harmonic representative which avoids the use of the computationally expensive derivatives of the Ricci-flat metric. The result follows from the following lemma.
\begin{lemma}\label{lemma:harmonicNannicini}
Let $X$ be Calabi--Yau with K\"ahler class $[\omega] \in H^{1,1}(X)\cap H^2(X)$ and $\alpha\in \Omega^{0,1}(T_X)$ is closed, then the following are equivalent:
\begin{enumerate}
\addtolength{\leftskip}{2pc}
	\item $\alpha$ is harmonic with respect to Ricci-flat metric in the K\"ahler class $[\omega]$.
	\item $\alpha$ is polarization preserving with respect to $[\omega]$ and $\partial\Omega(\alpha) = 0$.
\end{enumerate}
In statement~2, polarization preserving is in the sense of~\cite{tian:1987, nannicini:1986}.
\end{lemma}
\pf{%
That $1 \Rightarrow 2$ is proved in~\cite{nannicini:1986} and depends on Ricci-flatness of the metric. Conversely, to show that $2 \Rightarrow 1$, pick an open set $U\subset X$ and local coordinates $(z^1,\dots,z^n)$ on $U$ such that $\Omega = f(z) \rd z^1\wedge\dots \wedge \rd z^n$. Locally, we may expand $\partial\Omega(\alpha)$ as:
\begin{equation}
\begin{aligned}
	\partial\Omega(\alpha) &= \partial\left(\sum_{\mu=1}^n (-1)^{\mu-1}f(z)\,\alpha_{\overline{\nu}}^\mu\,\rd \overline{z}^\nu \wedge \widehat{\mu} \right)\,,\\
 &= \sum_{\mu=1}^n \frac{\partial}{\partial z^\mu}
 \big(f(z)\,\alpha^\mu_{\overline{\nu}}\big)
 \rd \overline{z}^\nu \wedge \rd z^1\wedge \dots \wedge 
 \rd z^n = 0\,,
\end{aligned}
\end{equation}
where $\widehat{\mu} := \rd z^1 \wedge \dots \wedge \widehat{\rd z^\mu} \wedge \dots \wedge \rd z^n$. Similarly, the polarization preserving condition, at the level of forms, can be written as:
\begin{gather}
g_{\mu\overline{\rho}}\,g^{\sigma\overline{\nu}}\,
\alpha_{\overline{\nu}}^\mu 
= \alpha_{\overline{\rho}}^\sigma\,.
\end{gather}
Finally, using the local Monge--Amp\`ere equation, $\det g = \kappa |f(z)|^2$, we obtain:
\begin{equation}
\begin{aligned}
	\frac{\partial}{\partial z^\mu}
 \big(g_{\sigma\overline{\nu}}\, g^{\mu\overline{\rho}}\, \alpha^{\sigma}_{\overline{\rho}}\,\det g\big)
 \rd \overline{z}^\nu 
 &= \kappa\frac{\partial}{\partial z^\mu}
\big(|f(z)|^2 g_{\sigma\overline{\nu}}\,g^{\mu\overline{\rho}}\,\alpha^{\sigma}_{\overline{\rho}}\big)\rd\overline{z}^\nu\,,\\
&=\kappa\overline{f(z)}\frac{\partial}{\partial z^\mu}
 \big( f(z)\,\alpha^\mu_{\overline{\nu}}\big)
 \rd\overline{z}^\nu 
 = 0\,.
\end{aligned}
\end{equation}
Equivalently, extending the results globally to $X$, we have: $\overline{\partial}^\dagger \alpha = 0$ and since $\overline{\partial}\alpha = 0$ by definition, we see that $\alpha$ is indeed harmonic.
}

\bibliographystyle{JHEP}
\bibliography{ref}

\end{document}